\pdfoutput=1
\documentclass[12pt,a4paper,titlepage]{article}
\usepackage[utf8]{inputenc}
\usepackage[top=50pt,bottom=50pt,left=68pt,right=66pt]{geometry}
\usepackage{amsmath}
\usepackage{booktabs}
\usepackage{amssymb}
\usepackage[title]{appendix}
\usepackage{mathrsfs}
\usepackage{float}
\usepackage{multirow}
\usepackage{graphicx,caption,subcaption}
\usepackage[space]{grffile}
\newcommand{\overbar}[1]{\mkern1.5mu\overline{\mkern-1.5mu#1\mkern-1.5mu}\mkern 1.5mu}

\begin{document}
\renewcommand{\arraystretch}{1.3}

\makeatletter
\def\@hangfrom#1{\setbox\@tempboxa\hbox{{#1}}%
      \hangindent 0pt
      \noindent\box\@tempboxa}
\makeatother


\def\un#1{\relax\ifmmode\@@underline#1\else
        $\@@underline{\hbox{#1}}$\relax\fi}


\let\under=\unt                 
\let\ced=\ce                    
\let\du=\du                     
\let\um=\Hu                     
\let\sll=\lp                    
\let\Sll=\Lp                    
\let\slo=\os                    
\let\Slo=\Os                    
\let\tie=\ta                    
\let\br=\ub                     


\def\a{\alpha}
\def\b{\beta}
\def\c{\chi}
\def\d{\delta}
\def\e{\epsilon}
\def\f{\phi}
\def\g{\gamma}
\def\h{\eta}
\def\i{\iota}
\def\j{\psi}
\def\k{\kappa}
\def\l{\lambda}
\def\m{\mu}
\def\n{\nu}
\def\o{\omega}
\def\p{\pi}
\def\q{\theta}
\def\r{\rho}
\def\s{\sigma}
\def\t{\tau}
\def\u{\upsilon}
\def\x{\xi}
\def\z{\zeta}
\def\D{\Delta}
\def\F{\Phi}
\def\G{\Gamma}
\def\J{\Psi}
\def\L{\Lambda}
\def\O{\Omega}
\def\P{\Pi}
\def\Q{\Theta}
\def\S{\Sigma}
\def\U{\Upsilon}
\def\X{\Xi}


\def\ve{\varepsilon}
\def\vf{\varphi}
\def\vr{\varrho}
\def\vs{\varsigma}
\def\vq{\vartheta}


\def\ca{{\cal A}}
\def\cb{{\cal B}}
\def\cc{{\cal C}}
\def\cd{{\cal D}}
\def\ce{{\cal E}}
\def\cf{{\cal F}}
\def\cg{{\cal G}}
\def\ch{{\cal H}}
\def\ci{{\cal I}}
\def\cj{{\cal J}}
\def\ck{{\cal K}}
\def\cl{{\cal L}}
\def\cm{{\cal M}}
\def\cn{{\cal N}}
\def\co{{\cal O}}
\def\cp{{\cal P}}
\def\cq{{\cal Q}}
\def\car{{\cal R}}
\def\cs{{\cal S}}
\def\ct{{\cal T}}
\def\cu{{\cal U}}
\def\cv{{\cal V}}
\def\cw{{\cal W}}
\def\cx{{\cal X}}
\def\cy{{\cal Y}}
\def\cz{{\cal Z}}


\def\Sc#1{{\hbox{\sc #1}}}      
\def\Sf#1{{\hbox{\sf #1}}}      



\def\slpa{\slash{\pa}}                            
\def\slin{\SLLash{\in}}                                   
\def\bo{{\raise-.3ex\hbox{\large$\Box$}}}               
\def\cbo{\Sc [}                                         
\def\pa{\partial}                                       
\def\de{\nabla}                                         
\def\dell{\bigtriangledown}                             
\def\su{\sum}                                           
\def\pr{\prod}                                          
\def\iff{\leftrightarrow}                               
\def\conj{{\hbox{\large *}}}                            
\def\ltap{\raisebox{-.4ex}{\rlap{$\sim$}} \raisebox{.4ex}{$<$}}   
\def\gtap{\raisebox{-.4ex}{\rlap{$\sim$}} \raisebox{.4ex}{$>$}}   
\def\TH{{\raise.2ex\hbox{$\displaystyle \bigodot$}\mskip-4.7mu \llap H \;}}
\def\face{{\raise.2ex\hbox{$\displaystyle \bigodot$}\mskip-2.2mu \llap {$\ddot
        \smile$}}}                                      
\def\dg{\sp\dagger}                                     
\def\ddg{\sp\ddagger}                                   

\font\tenex=cmex10 scaled 1200


\def\sp#1{{}^{#1}}                              
\def\sb#1{{}_{#1}}                              
\def\oldsl#1{\rlap/#1}                          
\def\slash#1{\rlap{\hbox{$\mskip 1 mu /$}}#1}      
\def\Slash#1{\rlap{\hbox{$\mskip 3 mu /$}}#1}      
\def\SLash#1{\rlap{\hbox{$\mskip 4.5 mu /$}}#1}    
\def\SLLash#1{\rlap{\hbox{$\mskip 6 mu /$}}#1}      
\def\PMMM#1{\rlap{\hbox{$\mskip 2 mu | $}}#1}   %
\def\PMM#1{\rlap{\hbox{$\mskip 4 mu ~ \mid $}}#1}       %
\def\Tilde#1{\widetilde{#1}}                    
\def\Hat#1{\widehat{#1}}                        
\def\Bar#1{\overline{#1}}                       
\def\sbar#1{\stackrel{*}{\Bar{#1}}}             
\def\bra#1{\left\langle #1\right|}              
\def\ket#1{\left| #1\right\rangle}              
\def\VEV#1{\left\langle #1\right\rangle}        
\def\abs#1{\left| #1\right|}                    
\def\leftrightarrowfill{$\mathsurround=0pt \mathord\leftarrow \mkern-6mu
        \cleaders\hbox{$\mkern-2mu \mathord- \mkern-2mu$}\hfill
        \mkern-6mu \mathord\rightarrow$}
\def\dvec#1{\vbox{\ialign{##\crcr
        \leftrightarrowfill\crcr\noalign{\kern-1pt\nointerlineskip}
        $\hfil\displaystyle{#1}\hfil$\crcr}}}           
\def\dt#1{{\buildrel {\hbox{\LARGE .}} \over {#1}}}     
\def\dtt#1{{\buildrel \bullet \over {#1}}}              
\def\der#1{{\pa \over \pa {#1}}}                
\def\fder#1{{\d \over \d {#1}}}                 


\def\frac#1#2{{\textstyle{#1\over\vphantom2\smash{\raise.20ex
        \hbox{$\scriptstyle{#2}$}}}}}                   
\def\half{\frac12}                                        
\def\sfrac#1#2{{\vphantom1\smash{\lower.5ex\hbox{\small$#1$}}\over
        \vphantom1\smash{\raise.4ex\hbox{\small$#2$}}}} 
\def\bfrac#1#2{{\vphantom1\smash{\lower.5ex\hbox{$#1$}}\over
        \vphantom1\smash{\raise.3ex\hbox{$#2$}}}}       
\def\afrac#1#2{{\vphantom1\smash{\lower.5ex\hbox{$#1$}}\over#2}}    
\def\partder#1#2{{\partial #1\over\partial #2}}   
\def\parvar#1#2{{\d #1\over \d #2}}               
\def\secder#1#2#3{{\partial^2 #1\over\partial #2 \partial #3}}  
\def\on#1#2{\mathop{\null#2}\limits^{#1}}               
\def\bvec#1{\on\leftarrow{#1}}                  
\def\oover#1{\on\circ{#1}}                              

\def\[{\lfloor{\hskip 0.35pt}\!\!\!\lceil}
\def\]{\rfloor{\hskip 0.35pt}\!\!\!\rceil}
\def\Lag{{\cal L}}
\def\du#1#2{_{#1}{}^{#2}}
\def\ud#1#2{^{#1}{}_{#2}}
\def\dud#1#2#3{_{#1}{}^{#2}{}_{#3}}
\def\udu#1#2#3{^{#1}{}_{#2}{}^{#3}}
\def\calD{{\cal D}}
\def\calM{{\cal M}}

\def\szet{{${\scriptstyle \b}$}}
\def\ulA{{\un A}}
\def\ulM{{\underline M}}
\def\cdm{{\Sc D}_{--}}
\def\cdp{{\Sc D}_{++}}
\def\vTheta{\check\Theta}
\def\fracm#1#2{\hbox{\large{${\frac{{#1}}{{#2}}}$}}}
\def\ha{{\fracmm12}}
\def\tr{{\rm tr}}
\def\Tr{{\rm Tr}}
\def\itrema{$\ddot{\scriptstyle 1}$}
\def\ula{{\underline a}} \def\ulb{{\underline b}} \def\ulc{{\underline c}}
\def\uld{{\underline d}} \def\ule{{\underline e}} \def\ulf{{\underline f}}
\def\ulg{{\underline g}}
\def\items#1{\\ \item{[#1]}}
\def\ul{\underline}
\def\un{\underline}
\def\fracmm#1#2{{{#1}\over{#2}}}
\def\footnotew#1{\footnote{\hsize=6.5in {#1}}}
\def\low#1{{\raise -3pt\hbox{${\hskip 0.75pt}\!_{#1}$}}}

\def\Dot#1{\buildrel{_{_{\hskip 0.01in}\bullet}}\over{#1}}
\def\dt#1{\Dot{#1}}

\def\DDot#1{\buildrel{_{_{\hskip 0.01in}\bullet\bullet}}\over{#1}}
\def\ddt#1{\DDot{#1}}

\def\DDDot#1{\buildrel{_{_{\hskip 0.01in}\bullet\bullet\bullet}}\over{#1}}
\def\dddt#1{\DDDot{#1}}

\def\DDDDot#1{\buildrel{_{_{\hskip 
0.01in}\bullet\bullet\bullet\bullet}}\over{#1}}
\def\ddddt#1{\DDDDot{#1}}

\def\Tilde#1{{\widetilde{#1}}\hskip 0.015in}
\def\Hat#1{\widehat{#1}}


\newskip\humongous \humongous=0pt plus 1000pt minus 1000pt
\def\caja{\mathsurround=0pt}
\def\eqalign#1{\,\vcenter{\openup2\jot \caja
        \ialign{\strut \hfil$\displaystyle{##}$&$
        \displaystyle{{}##}$\hfil\crcr#1\crcr}}\,}
\newif\ifdtup
\def\panorama{\global\dtuptrue \openup2\jot \caja
        \everycr{\noalign{\ifdtup \global\dtupfalse
        \vskip-\lineskiplimit \vskip\normallineskiplimit
        \else \penalty\interdisplaylinepenalty \fi}}}
\def\li#1{\panorama \tabskip=\humongous                         
        \halign to\displaywidth{\hfil$\displaystyle{##}$
        \tabskip=0pt&$\displaystyle{{}##}$\hfil
        \tabskip=\humongous&\llap{$##$}\tabskip=0pt
        \crcr#1\crcr}}
\def\eqalignnotwo#1{\panorama \tabskip=\humongous
        \halign to\displaywidth{\hfil$\displaystyle{##}$
        \tabskip=0pt&$\displaystyle{{}##}$
        \tabskip=0pt&$\displaystyle{{}##}$\hfil
        \tabskip=\humongous&\llap{$##$}\tabskip=0pt
        \crcr#1\crcr}}


\def\eV{\,{\rm eV}}
\def\keV{\,{\rm keV}}
\def\MeV{\,{\rm MeV}}
\def\GeV{\,{\rm GeV}}
\def\TeV{\,{\rm TeV}}
\def\sv{\left<\sigma v\right>}
\def\({\left(}
\def\){\right)}
\def\cm{{\,\rm cm}}
\def\K{{\,\rm K}}
\def\kpc{{\,\rm kpc}}
\def\beq{\begin{equation}}
\def\eeq{\end{equation}}
\def\bea{\begin{eqnarray}}
\def\eea{\end{eqnarray}}


\newcommand{\be}{\begin{equation}}
\newcommand{\ee}{\end{equation}}
\newcommand{\nbe}{\begin{equation*}}
\newcommand{\nee}{\end{equation*}}

\newcommand{\fr}{\frac}
\newcommand{\lb}{\label}

\thispagestyle{empty}

{\hbox to\hsize{
\vbox{\noindent August 2021 \hfill IPMU21-0034}
\noindent  \hfill }

\noindent
\vskip2.0cm
\begin{center}

{\large\bf Exploring the parameter space of modified supergravity \\ 
                for double inflation and primordial black hole formation}

\vglue.3in

Ryotaro Ishikawa~${}^{a}$ and Sergei V. Ketov~${}^{a,b,c}$
\vglue.3in

${}^a$~Department of Physics, Tokyo Metropolitan University\\
1-1 Minami-ohsawa, Hachioji-shi, Tokyo 192-0397, Japan \\
${}^b$~Research School of High-Energy Physics, Tomsk Polytechnic University\\
2a Lenin Avenue, Tomsk 634028, Russian Federation\\
${}^c$~Kavli Institute for the Physics and Mathematics of the Universe (WPI)
\\The University of Tokyo Institutes for Advanced Study, Kashiwa 277-8583, Japan\\
\vglue.1in

ishikawa-ryotaro@ed.tmu.ac.jp, ketov@tmu.ac.jp
\end{center}

\vglue.3in

\begin{center}
{\Large\bf Abstract}  

\end{center}

We study the parameter space of the effective (with two scalars) models of cosmological inflation and primordial black hole (PBH) formation in the modified $(R+R^2)$ supergravity. Our models describe double inflation, whose first stage is driven by Starobinsky's scalaron coming from the $R^2$ gravity, and whose second stage is driven by another scalar belonging to the supergravity multiplet. The ultra-slow-roll regime between the two stages leads a large peak (enhancement) in the power spectrum of scalar perturbations, which results in efficient PBH formation. Both inflation and PBH formation are generic in our models, while those PBH can account for a significant part or the whole of dark matter. Some of the earlier proposed models in the same class are in tension (over $3\s$) with the observed value of the scalar tilt $n_s$, so that we study more general models with more parameters, and investigate the dependence of the cosmological tilts $(n_s,r)$ and the scalar power spectrum enhancement upon the parameters. The PBH masses and their density fraction (as part of dark matter) are also calculated. A good agreement (between $2\s$ and $3\s$) with the observed value of $n_s$ requires fine tuning of the parameters, and it is only realized in the so-called $\d$-models. Our models offer the (super)gravitational origin of inflation, PBH and dark matter together, and may be confirmed or falsified by future precision measurements of the cosmic microwave background radiation and PBH-induced gravitational waves.

\vglue.1in
\noindent 

\newpage

\section{Introduction}

Symmetry principles play the very important role in fundamental physics, especially well beyond the Standard Model 
 (SM) of elementary particles, where they provide advanced guidance to the phenomenological model building. Supersymmetry (SUSY) is one of such fundamental symmetry principles that was proposed many years ago as the nontrivial extension of the Poincar\'e symmetry. No SUSY signature at the large hadron collider (LHC)  rules out low-scale SUSY but does not exclude high-scale SUSY well beyond 10 TeV. Local SUSY implies the invariance under general coordinate transformations, so that supergravity as the theory of local SUSY is more fundamental than general relativity. The scale of cosmological inflation is expected at $10^{10}$ TeV that is well beyond the electro-weak scale, where quantum gravity corrections are expected to be significant also. Supergravity can serve as a {\it bridge between classical and quantum gravity}, which makes it indispensable at the super-high energy scales close to the GUT scale, see e.g., Refs.~\cite{Ellis:2020lnc,Ketov:2019mfc} for the recent reviews.

The standard (concordance) model of cosmology (SCM) assumes cold dark matter (CDM) and the positive cosmological constant as the dark energy (DE). However, it does not tell us about physics behind CDM and DE.
It is often assumed that CDM is composed of unknown (electrically neutral and stable or meta-stable) particles,
though it is also possible that part or the whole CDM may also be composed of primordial black holes (PBH) formed in the early Universe during inflation \cite{Novikov:1967tw,Hawking:1971ei,Carr:1974nx,Dolgov:1992pu,Barrow:1992hq}. 
We adopt the second possibility in this paper (PBH as DM), see also Refs.~\cite{Carr:2003bj,Sasaki:2018dmp,Carr:2020gox,Carr:2020xqk} for the recent reviews.

There is the well-known problem with the cosmological constant because its interpretation as the vacuum energy  contradicts quantum theory. A positive cosmological constant implies a de Sitter vacuum that
is incompatible with the S-matrix approach and, therefore, is inconsistent with quantum gravity  
\cite{Dvali:2020etd}. In particular, de Sitter vacua are believed to be forbidden in string theory  also \cite{Ooguri:2018wrx}.

Yet another well known indication on the necessity to go beyond the SCM is the Hubble tension, namely, a  disagreement between local measurements of Hubble parameter from supernova and lensing, on the one side, and its value inferred from the SCM fit to the cosmic microwave background (CMB) radiation 
\cite{Freedman:2017yms}.

To avoid (and mimic) a positive cosmological constant in the classical description of gravity, one can employ  the modified $f(R)$ gravity theory, where DE can be effectively described by a non-trivial function $f(R)$  of the (Ricci) scalar curvature in the gravitational action, in agreement with gravitational tests in Solar system \cite{Hu:2007nk,Appleby:2007vb,Starobinsky:2007hu}, see Ref.~\cite{DeFelice:2010aj} for a comprehensive review.

Cosmological inflation in the early Universe can also be successfully described in the  modified $f(R)$ gravity by the celebrated Starobinsky model of  $(R+R^2)$ gravity \cite{Starobinsky:1980te}, see also  Ref.~\cite{Ketov:2019toi} for a recent review of Starobinsky's inflation. The Starobinsky inflation model is more than the best fit to the current CMB observations. It is geometrical because it relies on gravitational interactions, has no free parameters, and is driven by the scale-invariant $R^2$ term in the action. Hence, Starobinsky's inflaton (called scalaron) has the clear gravitational origin as the physical degree of freedom in the higher-derivative gravity, while it can be identified with the Nambu-Goldstone boson related to spontaneous breaking of the scale invariance \cite{Ketov:2012jt}.

However, the Starobinsky model has limited applicability because the CMB data is just a small window into high-energy physics, while there are no observational constraints for the scales both well beyond and well under the (expected) inflationary scale of $10^{13}$ GeV, when being measured by the Hubble parameter. There is also no good reason for the absence of other scalars during inflation. Therefore, it is of importance to figure out possible interactions of scalaron with other particles before and during reheating. Specifying those scalaron interactions can be done either by a trial-and-error procedure or by postulating some fundamental (symmetry) principles. In this paper we assume Supersymmetry (SUSY) as the guiding fundamental symmetry principle, which means local SUSY or supergravity in the context of gravity.  Despite the current absence of experimental confirmation, SUSY is still one of the leading candidates for new physics beyond the SM. Of course, SUSY has to be (spontaneously) broken at the energies available for observations, see e.g., Refs.~\cite{Aldabergenov:2017hvp,Antoniadis:2018wjx} for specific mechanisms of spontaneous SUSY breaking after inflation in supergravity.

Therefore, the natural questions arise: (i) whether the modified supergravity can successfully describe inflation
and PBH formation in the early Universe, and (ii) how much fine tuning is needed for those purposes?

There exist many supergravity extensions of the Starobinsky inflation, see e.g., Ref.~\cite{Ketov:2019toi} for a brief review of the existing approaches to embedding the Starobinsky model into supergravity, and   Ref.~\cite{Ketov:2019rzg} for inclusion of DM and DE too. Supergravity generically leads to {\it multi-field} inflation with several scalar fields involved, whereas the Starobinsky inflation model is an example of {\it single-field} inflation. Most of the known embeddings of inflation into supergravity 
have a single inflaton and suppress other scalars. In this paper we employ the generalized (or modified) Starobinsky-type supergravity \cite{Ketov:2012jt,Ketov:2019rzg,Ketov:2010qz,Ketov:2013dfa,Addazi:2017rkc} in order to describe multi-field inflation. We minimize the particle spectrum by keeping all fields in the {\it same} (irreducible) supergravity multiplet, while their interactions are severely restricted  by local SUSY. We give a supergravity realization of two-field double inflation with a sharp turn of the inflationary trajectory, where one scalar is given by Starobinsky's inflaton and another scalar is its scalar superpartner. We show that it leads to PBH generation and DM genesis after Starobinsky inflation. Our models in this paper do not have a cosmological constant  and have Minkowski vacua, without addressing DE. However, they allow further extensions that may lead to a viable DE description also, which is beyond the scope of this investigation.

Our paper is organized as follows. In Sec.~2 we introduce our setup by defining the modified Starobinsky-type supergravity and deriving the effective two-field inflationary models, which coincides with the setup adopted in Refs.~\cite{Aldabergenov:2020bpt,Ketov:2021fww}. In Sec.~3 we review the two special cases already studied in Ref.~\cite{Aldabergenov:2020bpt}, having one parameter less, and called the $\gamma$ and $\delta$ models, respectively. The main purpose of this investigation is to study the models where both parameters $\gamma$ and $\delta$ are non-vanishing, as well as even more general models beyond Ref.~\cite{Aldabergenov:2020bpt}. The significance of the new cases stems from the observation that the $\gamma$-models with one less parameter are  in tension (over $3\s$) with the observed value of the CMB scalar spectral index $n_s$. It was conjectured in Ref.~\cite{Aldabergenov:2020bpt} that one may get a much better agreement in more general models with both non-vanishing  $\gamma$ and $\delta$ parameters, or by adding more parameters. We present our new findings in Sec.~4 that contains our main results. Our Conclusion is given by Sec.~5.

The modified (Starobinsky) supergravity is used to generate the scalar potential, instead of postulating it  
{\it ad hoc}. Once the scalar potential is derived, the standard techniques are applied to compute the power spectra
of scalar and tensor perturbations and the inflationary observables. In the case of two-field inflation, those techniques are described at length in Refs.~\cite{Aldabergenov:2020bpt,Gundhi:2018wyz,Gundhi:2020kzm}, so that we skip their  details here and focus on our new results.

\section {Modified Starobinsky supergravity and the effective two-field models of inflation and PBH formation}

We use the standard notation of Ref.~\cite{Wess:1992cp} in curved superspace of the (old-minimal) supergravity. It guarantees that our starting actions are manifestly $N=1$ locally supersymmetric in four spacetime dimensions. As regards the supersymmetry transformation laws of the field components, see Ref.~\cite{Wess:1992cp}.

The chiral superspace Lagrangian of the chiral superfields $\mathbf{\Phi}^i$ coupled to supergravity reads (we take the reduced Planck mass $M_{\rm Pl}=1$ for simplicity)
\begin{equation}
    {\cal L}=\int d^2\Theta 2{\cal E}\left[\frac{3}{8}(\overbar{\cal D}^2-8{\cal R})e^{-K(\mathbf{\Phi}^i,\overbar{\mathbf{\Phi}}^i)/3}+W(\mathbf{\Phi}^i)\right]+{\rm h.c.}~,\label{App_L_super}
\end{equation}
where we have introduced the chiral density superfield $\cal E$, the chiral curvature superfield
 $\cal R$, the superspace covariant derivatives  ${\cal D}_\alpha,\overbar{\cal D}_{\dot{\alpha}}$~, with 
 ${\cal D}^2\equiv{\cal D}^\alpha{\cal D}_\alpha$ and $\overbar{\cal D}^2\equiv\overbar{\cal D}_{\dot{\alpha}}\overbar{\cal D}^{\dot{\alpha}}$, the K\"ahler potential $K(\Phi^i,\overbar{\Phi}^i)$  and the superpotential $W(\Phi^i)$. The  $K$ and $W$ define the model and uniquely determine its scalar sector.
   
Eliminating the auxiliary fields and going to Einstein frame via field redefinitions yield the bosonic part of the Lagrangian  (\ref{App_L_super}) in the form
\begin{equation}
    e^{-1}{\cal L}=\frac{1}{2}R-K_{i\bar{j}}\partial_m\Phi^i\partial^m\overbar{\Phi}^j-e^K\left(K^{i\bar{j}}D_iW D_{\bar{j}}\overbar{W}-3|W|^2\right)~,\label{App_L}
\end{equation}
where we have used the same notation for the chiral superfields and their leading (scalar) field components, together with 
\begin{gather}
    K_{i\bar{j}}\equiv\fracmm{\partial^2 K}{\partial\Phi^i\partial\overbar{\Phi}^j}~,\quad K^{i\bar{j}}\equiv K^{-1}_{i\bar{j}}~,\quad {\rm and}\quad D_iW\equiv\fracmm{\partial W}{\partial\Phi^i}+W\fracmm{\partial K}{\partial\Phi^i}~~.
\end{gather}

The expansions of the supergravity chiral superfields $\cal E$ and $\cal R$ with respect to the chiral 
anti-commuting coordinates $\Theta_{\a}$ define their field components as follows:
\begin{eqnarray}
    2{\cal E}&=&e\left[1+i\Theta\sigma^m\overbar{\psi}_m+\Theta^2(6\overbar{X}-\overbar{\psi}_m\overbar{\sigma}^{mn}\overbar{\psi}_n)\right]~,\label{E_expansion}\\
    {\cal R}&=&X+\Theta\left(-\frac{1}{6}\sigma^m\overbar{\sigma}^n\psi_{mn}-i\sigma^m\overbar{\psi}_mX-\frac{i}{6}\psi_mb^m\right)+\nonumber\\
    &~&+\Theta^2\left(-\frac{1}{12}R-\frac{i}{6}\overbar{\psi}^m\overbar{\sigma}^n\psi_{mn}-4X\overbar{X}-\frac{1}{18}b_mb^m+\frac{i}{6}\nabla_mb^m+\right.\nonumber\\
    &~&+\left.\frac{1}{2}\overbar{\psi}_m\overbar{\psi}^mX+\frac{1}{12}\psi_m\sigma^m\overbar{\psi}_nb^n-\frac{1}{48}\varepsilon^{abcd}(\overbar{\psi}_a\overbar{\sigma}_b\psi_{cd}+\psi_a\sigma_b\overbar{\psi}_{cd})\right)~,\label{R_expansion}
\end{eqnarray}
where $e\equiv{\rm det}(e^a_m)$, $\psi_{mn}\equiv\tilde{D}_m\psi_n-\tilde{D}_n\psi_m$ and 
$\tilde{D}_m\psi_n\equiv(\partial_m+\omega_m)\psi_n$.  \

The supergravity multiplet includes vierbein  $e^a_m$, gravitino $\psi_m$, the real vector field $b_m$ and the complex scalar $X$. The chiral superfield ${\cal E}$ can be seen as the SUSY extension of the spacetime density $e=\sqrt{-g}$, and the chiral superfield ${\cal R}$ can be seen as the SUSY extension of the (Ricci) scalar curvature $R$. The real vector $b_m$ and the complex scalar $X$ are known in the supergravity literature  as the (old-minimal) set of the "auxiliary" fields needed to complete the supergravity multiplet off-shell. In the {\it higher-derivative} modified supergravity (see below), those "auxiliary" fields become the physical ones, i.e. they are dynamical or propagating.

The modified Starobinsky-type supergravity action in curved (full) superspace is defined by  \cite{Cecotti:1987sa,Gates:2009hu}
\begin{equation}
\label{A}
S=\int d^{4}x d^{4}\theta E^{-1}N(\mathcal{R},\bar{\mathcal{R}})+\left[\int d^{4}x d^{2}\Theta 2\mathcal{E}{\cal F}(\mathcal{R})+h.c    \right]~~.
\end{equation}
The equivalent action in the curved chiral superspace reads
\begin{equation}
    {\cal L}=\int d^2\Theta 2{\cal E}\left[-\frac{1}{8}(\overbar{\cal D}^2-8{\cal R})N({\cal R},\overbar{\cal R})+{\cal F}({\cal R})\right]+{\rm h.c.} \label{L_master}
\end{equation}
This action is governed by two arbitrary potentials $N$ and ${\cal F}$ that are similar to $K$ and $F$
in Eq.~(\ref{App_L_super}), respectively. Actually, the ${\cal F}$-type term in Eq.~(\ref{A})
can be included into the $N$-type term, unless the former is a constant. However, we prefer to distinguish the two structures because they represent two different deformations of the simplest supergravity action. It is worth mentioning that, unlike the chiral and anti-chiral (matter) superfields in Eq.~(\ref{App_L_super}), the arguments of the functions $N$ and ${\cal F}$ are the {\it supergravity} superfields,  ${\cal R}$ and $\overbar{\cal R}$, that are not unconstrained chiral superfields but satisfy the supergravity constraints in curved superspace. The chiral superfield  $\mathcal{R}$ has the scalar curvature $R$  as its field component at $\Theta^2$, so that {\it no higher powers}  of $R$ (beyond the linear and quadratic terms) appear in the action (\ref{A}). Therefore, supergravity apparently distinguishes the $(R+R^2)$ model amongst all the $f(R)$ gravity models! The higher powers of $R$ can, nevertheless, appear when we allow the superspace derivatives of the supergravity superfields ${\cal R}$ and $\overbar{\cal R}$ in the arguments of $N$ and ${\cal F}$. However, it would lead to the appearance of the spacetime derivatives of the scalar curvature in  the Lagrangian and more physical scalars.

The action (\ref{A}) can be transformed into the (dual) standard matter-coupled supergravity action of the type (\ref{App_L_super}) in terms of {\it two} chiral matter superfields  $T$ and $S$ in the manifestly supersymmetric way, whose K\"ahler potential and the superpotential are related to the input potentials $N$ and $F$ as follows~\cite{Cecotti:1987sa,Gates:2009hu,Aldabergenov:2020yok}:
\begin{equation} K= -3\log \left[T+\overbar{T}-\frac{1}{3}N(S,\overbar{S})\right] \quad {\rm and}\quad  
W=3ST+{\cal F}(S)~.\label{duals}
\end{equation}
This K\"ahler potential is an example of the {\it no-scale} supergravity that naturally arises in compactifications of 
heterotic strings \cite{Ellis:2013nka}. Unlike the non-SUSY duality between $f(R)$ gravity and scalar-tensor gravity, there is no inverse transformation that could relate any choice of $K$ and $W$ to the potentials $N$ and ${\cal F}$ of the generalized Starobinsky-type supergravity. The particular form (\ref{duals}) distinguishes  the induced K\"ahler potentials $K$ and the superpotentials  $W$ of the supergravitational origin (from the modified supergravity).

Let us expand the functions $N$ and $\cal F$ in Taylor series and keep a few leading terms as follows:
\begin{gather}
    N=\fracmm{12}{M^2}|{\cal R}|^2-\fracmm{72}{M^4}\zeta|{\cal R}|^4-\fracmm{768}{M^6}\gamma|{\cal R}|^6~~,
    \label{N_choice2}\\
    {\cal F}=-3{\cal R}+\fracmm{3\sqrt{6}}{M}\delta {\cal R}^2~~,\label{N_F_choice2}
\end{gather}
where we have introduced three {\it dimensionless} parameters $\zeta$, $\gamma$ and $\delta$, and the Starobinsky mass $M$. This supergravity model reduces to the standard (pure) supergravity in the very special case of $N=0$ and ${\cal F}=-3{\cal R}$. In particular, the first term in Eq.~(\ref{N_F_choice2}) leads to the Einstein-Hilbert 
term $\ha eR$ in the Lagrangian, according to Eqs.~(\ref{E_expansion}) and (\ref{R_expansion}).  Similarly, the first term in the $N$-potential (\ref{N_choice2}) is needed to generate the $R^2$ term in the Lagrangian with the Starobinsky mass parameter $M$. We recall that the value of $M\sim 10^{-5}$ is fixed by CMB observations. The second term in (\ref{N_choice2}) is needed for stabilization of the 
Starobinsky supergravity, see Ref.~\cite{Addazi:2017rkc} for details. The other $\gamma$ and $\delta$ terms
represent further deformations of that Starobinsky-type modified supergravity \cite{Aldabergenov:2020bpt}, whose impact on PBH production is reviewed in the next Section.

Let us also ignore the vector field $b_m$ and the angular part of the scalar field ${\cal R}|=X$ for simplicity (the scalar
potenial is flat in the angular direction), and rescale the real $X$ as 
\begin{equation} \label{setX}
X=\fracmm{M\sigma}{\sqrt{24}}~.
\end{equation}

 The dual matter-coupled Einstein supergravity is described by  Eq.~(\ref{duals}) with the functions
\begin{align}
    N(S,\overbar{S})&=3\left(|S|^2-\frac{3}{2}\zeta|S|^4-4\gamma|S|^6\right)~,\\
    {\cal F}(S)&=3MS\left(\frac{\sqrt{6}}{4}\delta S-\frac{1}{2}\right)~,
\end{align}
after rescaling ${\cal R}=MS/2$. Equation \eqref{setX} implies $S=\sigma/\sqrt{6}$. The canonical scalaron 
$\varphi$ in the dual picture is given by
\begin{equation}
    e^{\sqrt{\frac{2}{3}}\varphi}=T+\overbar{T}-\frac{1}{3}N(S,\overbar{S})~.
\end{equation}

It is straightforward (after a long calculation) to get the scalar part of the Lagrangian in Einstein frame. It takes
the form of two scalar fields coupled to gravity as follows 
 \cite{Aldabergenov:2020bpt}:
\begin{equation}
    e^{-1}{\cal L}=\fracmm{1}{2}R-\fracmm{1}{2}(\partial\varphi)^2-\fracmm{3M^2}{2}Be^{-\sqrt{\frac{2}{3}}\varphi}(\partial\sigma)^2-\fracmm{1}{4B}\left(1-Ae^{-\sqrt{\frac{2}{3}}\varphi}\right)^2-e^{-2\sqrt{\frac{2}{3}}\varphi}U~,\label{L_varphi2}
\end{equation}
where the functions $A,B,U$ are given by
\begin{align}
    A&=1-\delta\sigma+\fracmm{1}{6}\sigma^2-\fracmm{11}{24}\zeta\sigma^4-\fracmm{29}{54}\gamma\sigma^6~,\nonumber\\
    B&=\fracmm{1}{3M^2}(1-\zeta\sigma^2-\gamma\sigma^4)~,\label{ABU_tilde}\\
    U&=\fracmm{M^2}{2}\sigma^2\left(1+\fracmm{1}{2}\delta\sigma-\fracmm{1}{6}\sigma^2+\frac{3}{8}\zeta\sigma^4+\frac{25}{54}\gamma\sigma^6\right)~.\nonumber 
\end{align}

Both scalars, scalaron $\varphi$ and extra scalar $\s$, have the supergravitational origin and are related via
local SUSY transformations by construction. Their kinetic terms and the scalar potential are determined by Eqs.~(\ref{L_varphi2}) and (\ref{ABU_tilde}). In particular,  they have the same masses. One may worry about $\s$ to become a ghost due to the negative signs at some terms in the second equation  (\ref{ABU_tilde}). Actually, the infinite wall in the scalar potential prevents $\sigma$ from obtaining the values leading to the wrong sign of its kinetic term. Thus, there are no ghosts.

It is also straightforward (after a long calculation) to derive the equations of motion of our two-field model in the FLRW universe, when keeping only time dependence of the fields. One finds  \cite{Aldabergenov:2020bpt}
\begin{align}
    &\ddot\varphi+3H\dot{\varphi}+\fracmm{1}{\sqrt{6}}(1-\zeta\sigma^2-\gamma\sigma^4)e^{-\sqrt{\frac{2}{3}}\varphi}\dot{\sigma}^2+\partial_\varphi V=0~,\label{KG1_gamma}\\
    &\ddot\sigma+3H\dot{\sigma}-\fracmm{\zeta\sigma+2\gamma\sigma^3}{1-\zeta\sigma^2-\gamma\sigma^4}\dot{\sigma}^2-\sqrt{\fracmm{2}{3}}\dot{\varphi}\dot{\sigma}+\fracmm{e^{\sqrt{\frac{2}{3}}\varphi}}{1-\zeta\sigma^2-\gamma\sigma^4}\partial_\sigma V=0~,\label{KG2_gamma}\\
    &\frac{1}{2}\dot{\varphi}^2+\frac{1}{2}(1-\zeta\sigma^2-\gamma\sigma^4)e^{-\sqrt{\frac{2}{3}}\varphi}
    \dot{\sigma}^2+\dot{H}=0~,\label{Fried1_gamma}
\end{align}
where the last equation describes the kinetic energy balance. These equations are to be supplemented by the Friedman equation for completeness,
\begin{equation} \label{Fried2_gamma}
V=3H^2 +\dot{H}~,
\end{equation}
where the full scalar potential $V$ is given by the last two terms in Eq.~(\ref{L_varphi2}).

The main cosmological parameters of inflation are given by the scalar tilt $n_s$ and the tensor-to-scalar ratio $r$,
whose observational values are constrained by the Planck measurements of CMB as \cite{Planck:2018jri}
\begin{equation} \label{PlanckCMB}
n_s=0.9649\pm 0.0042  \quad ({\rm 68\% CL}) \qquad {\rm and} \qquad  r < 0.056 \quad ({\rm 95\% CL})~.
\end{equation}

\section{Two special cases}

In this Section we review the results of Ref.~\cite{Aldabergenov:2020bpt} for two special families of our models with one less free parameter, and then compare them. It serves as the pre-requisite for our new investigation in the next Section. Those special cases are the one with  $\delta=0$ (called the $\gamma$ models) and another one with $\gamma=0$ (called the $\delta$ models). Simultaneously, this Section introduces our methods of investigation of PBH production, which are used in the next Section also. More technical details can be found in Refs.~\cite{Gundhi:2018wyz,Gundhi:2020kzm}.

\subsection{The $\gamma$ models}

Let us choose the case with the parameters $\gamma=1$ and $\zeta=-1.7774$, as a representative of the $\g$ models ($\delta=0$) The scalar potential for $\varphi\gg 1$ in Fig.~\ref{V_3d_gamma}  has two valleys at $\sigma\neq 0$, and a single Minkowski minimum at $\sigma=\varphi=0$. The first slow-roll (SR) Starobinsky-like inflation goes along either of the valleys. The inflationary trajectory has a sharp turn by passing through one of the near-inflection points followed by the second inflation stage toward the Minkowski minimum. In the very short so-called "ultra-slow-roll" (USR) regime between the two stages of inflation, the scalar field(s) roll down the potential faster than in the SR regime \cite{Motohashi:2014ppa}.

\begin{figure}
\centering
\begin{subfigure}{.49\textwidth}
  \centering
  \includegraphics[width=1\linewidth]{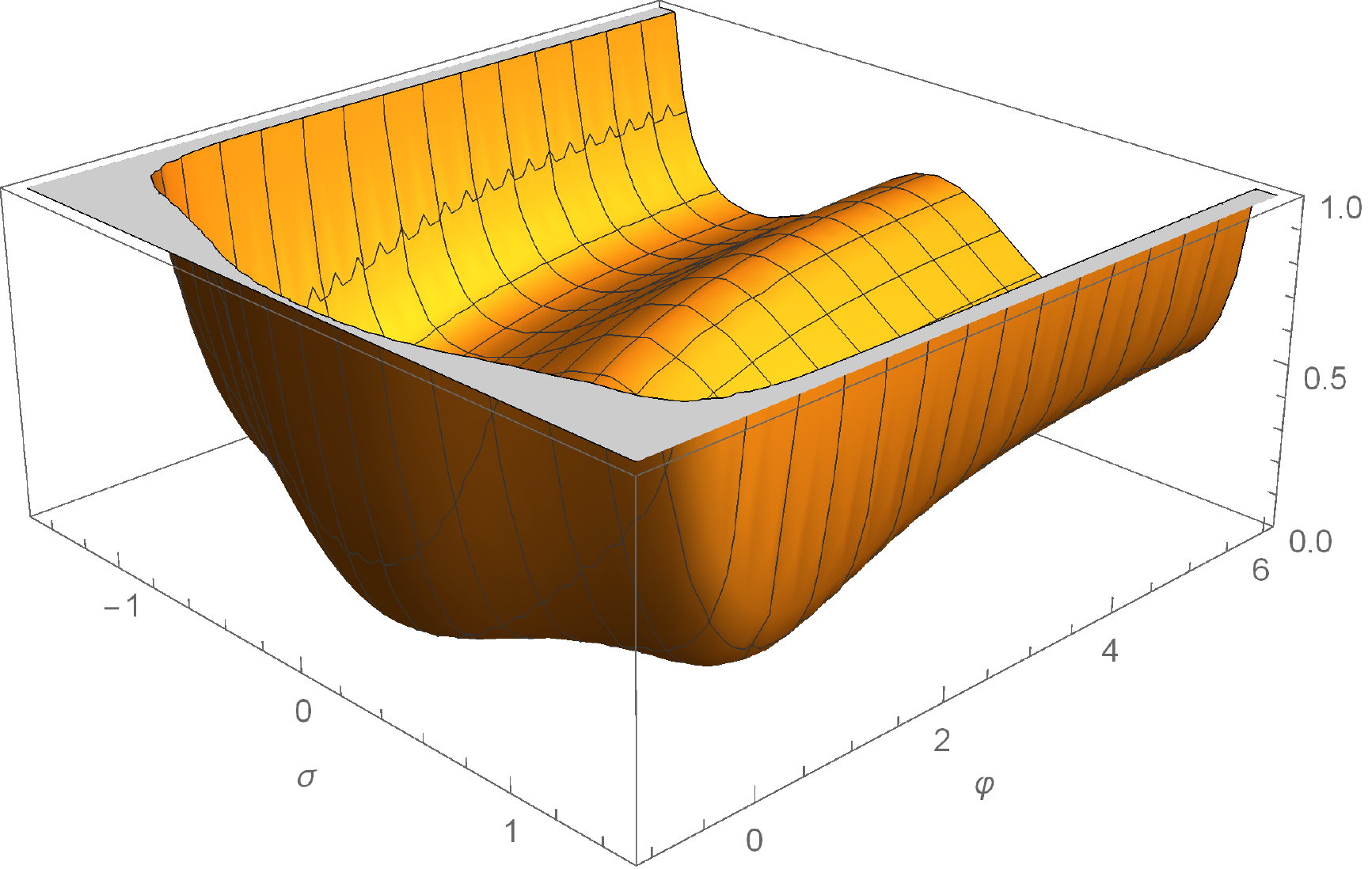}
  \label{V_3d1_gamma}
\end{subfigure}
\begin{subfigure}{.49\textwidth}
  \centering
  \includegraphics[width=.85\linewidth]{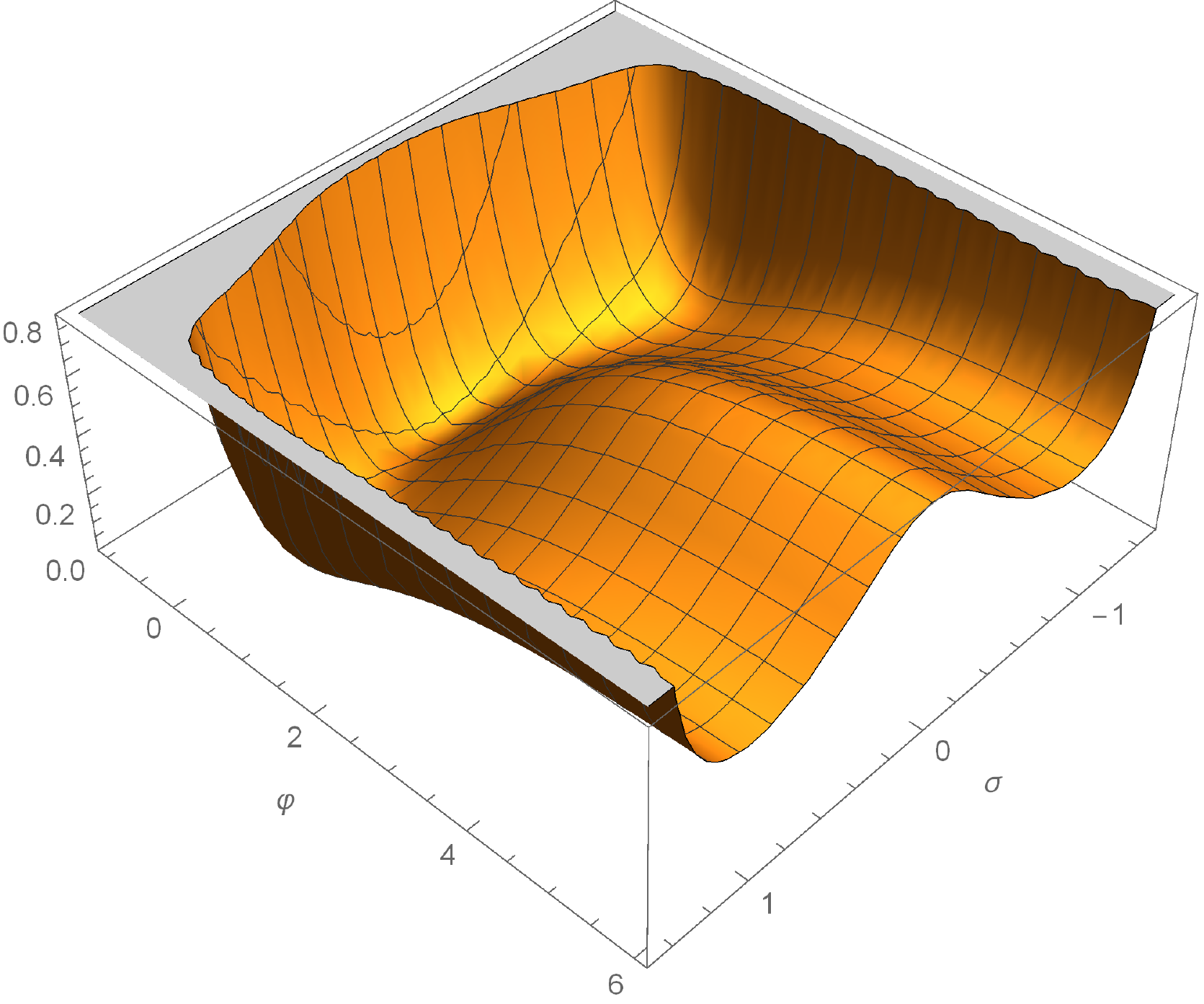}
  \label{V_3d2_gamma}
\end{subfigure}
\captionsetup{width=.9\linewidth}
\caption{The scalar potential $V/M^2$ for $\delta=0$, $\gamma=1$ and $\zeta=-1.7774$.}
\label{V_3d_gamma}
\end{figure}

The numerical solutions to the equations of motion are plotted in Fig.~\ref{fsV_sol_gamma}. The total number of e-folds is set to $\Delta N=60$, and the end of the first stage of inflation is defined by the time when the SR parameter $\eta$ first reaches $1$, see Fig.~\ref{en_sol_gamma}.  The USR period $\epsilon_{\rm USR}\ll\epsilon_{\rm SR}$ in Fig.~\ref{en_sol_gamma} between the two stages of inflation leads to a significant {\it enhancement} of the scalar power spectrum, see Fig.~\ref{Pk_int_gamma}. Inflation ends when $\epsilon=1$. The first stage lasts $\Delta N_1\approx 50$ e-folds, whereas the second stage lasts for  $\Delta N_2\approx 10$.  The length of the second stage is controlled by the parameter $\zeta$ at a given $\gamma$.

\begin{figure}
\centering
\begin{subfigure}{.49\textwidth}
  \centering
  \includegraphics[width=0.85\linewidth]{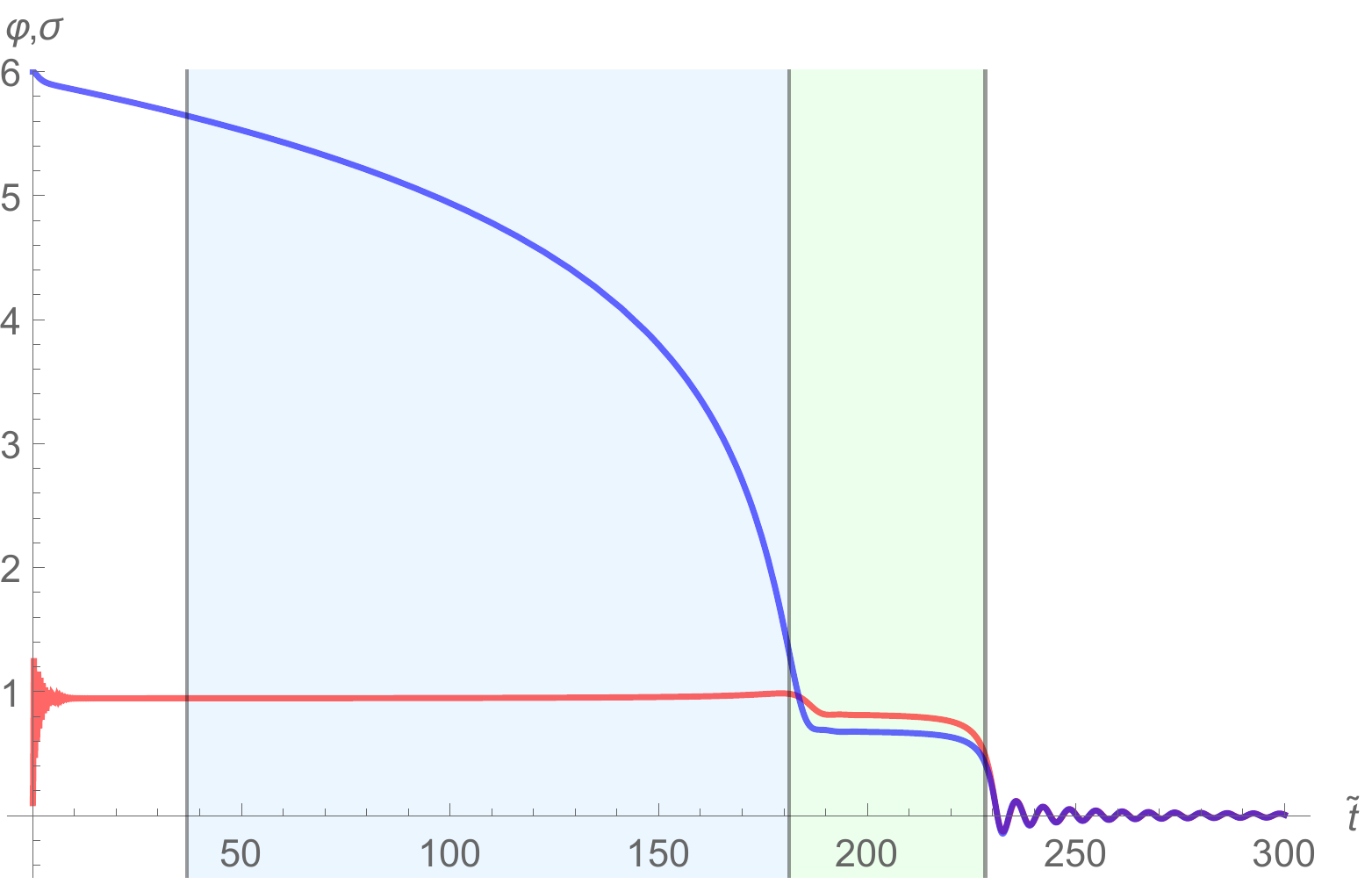}
  \caption{}
  \label{fs_sol_gamma}
\end{subfigure}
\begin{subfigure}{.49\textwidth}
  \centering
  \includegraphics[width=.75\linewidth]{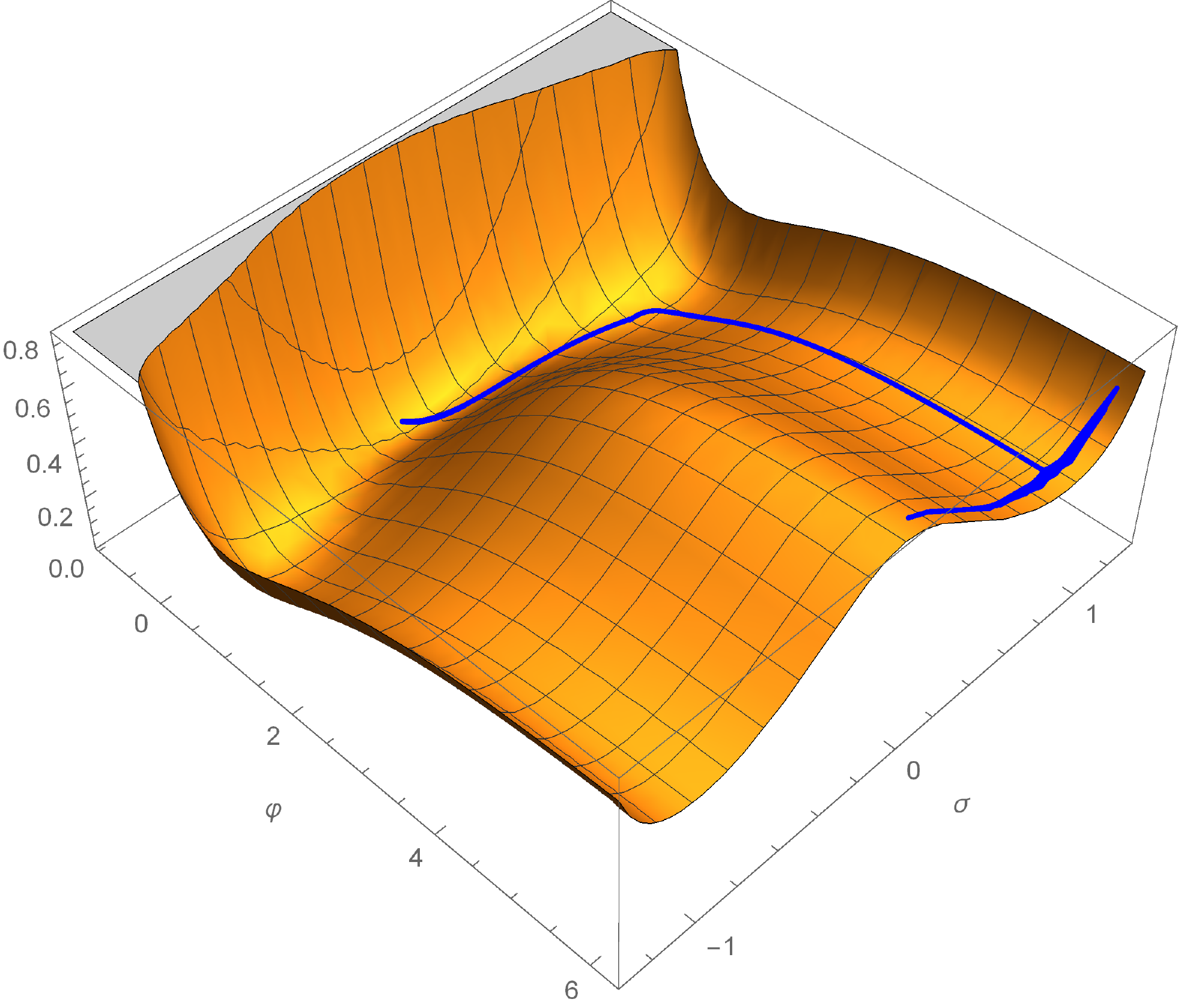}
  \caption{}
  \label{V_sol_gamma}
\end{subfigure}
\begin{subfigure}{.32\textwidth}
  \centering
  \includegraphics[width=.9\linewidth]{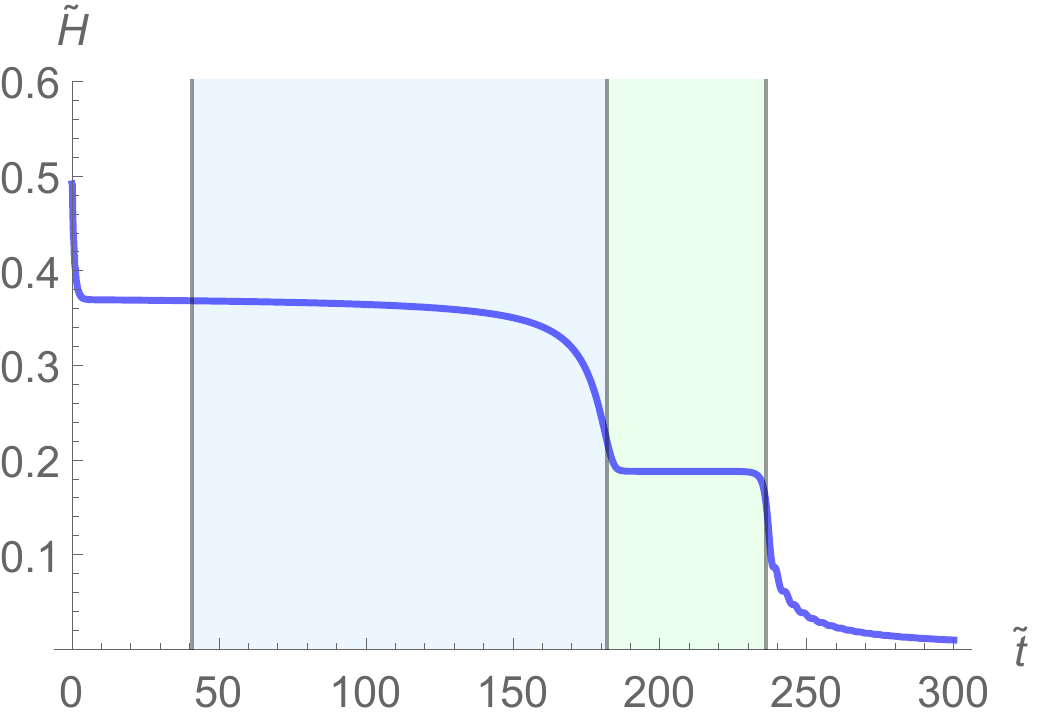}
  \caption{}
  \label{H_sol_gamma}
\end{subfigure}
\begin{subfigure}{.32\textwidth}
  \centering
  \includegraphics[width=.9\linewidth]{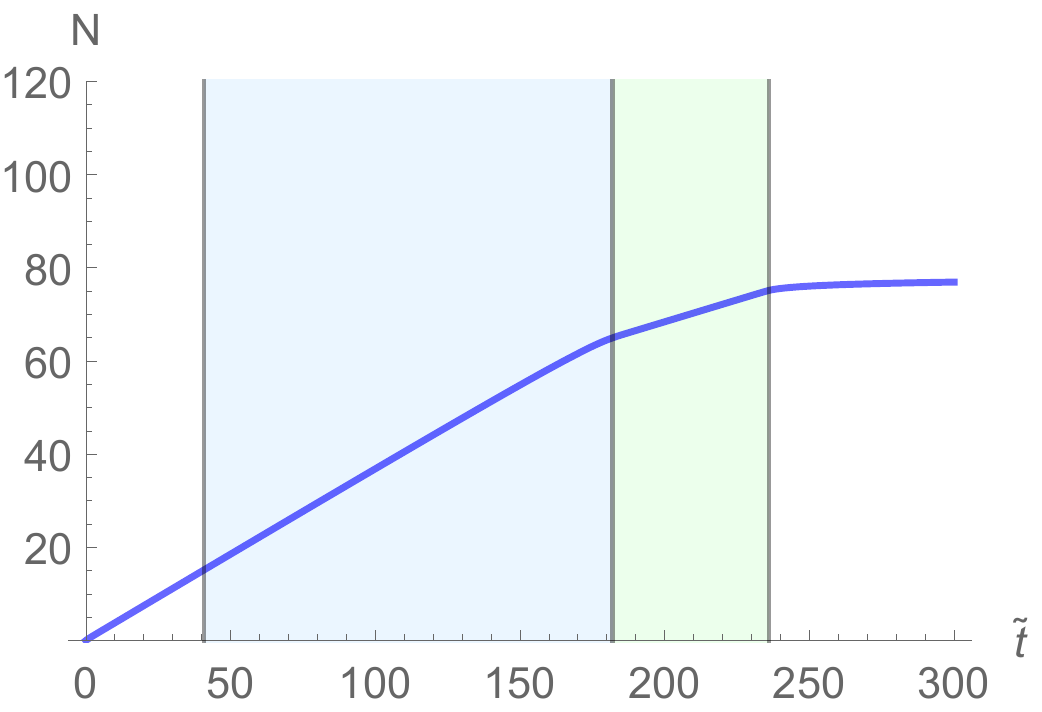}
  \caption{}
  \label{N_sol_gamma}
\end{subfigure}
\begin{subfigure}{.32\textwidth}
\centering
\includegraphics[width=.95\linewidth]{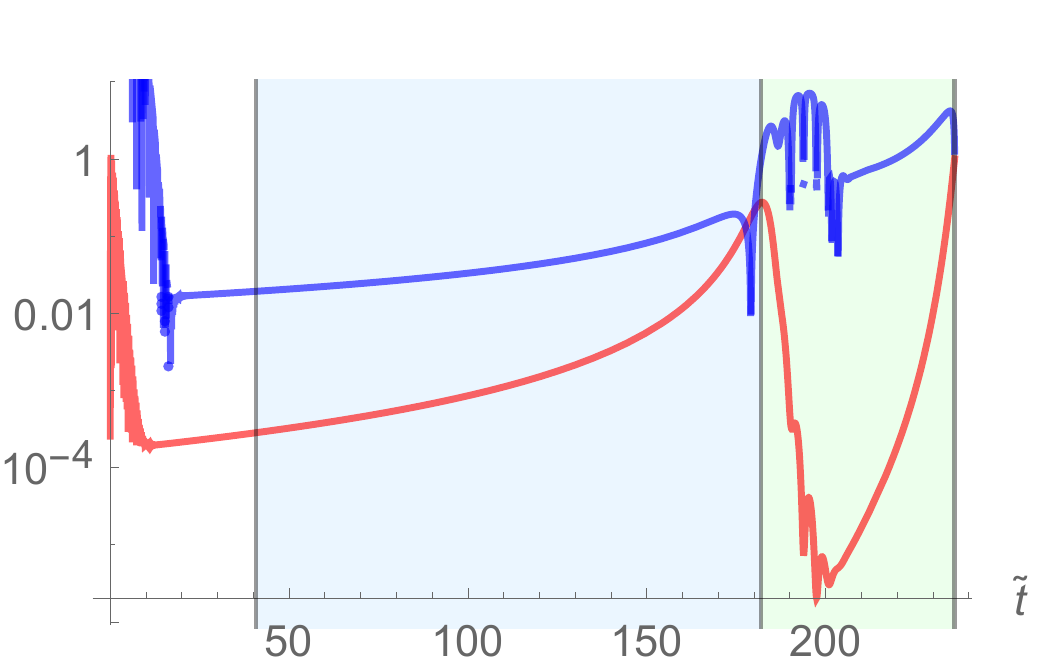}
\captionsetup{width=.9\linewidth}
\caption{}
\label{en_sol_gamma}
\end{subfigure}
\captionsetup{width=.9\linewidth}
\caption{(a) The solution to the field equations \eqref{KG1_gamma} and \eqref{KG2_gamma} with the initial conditions $\varphi(0)=6,\sigma(0)=0.1$, the vanishing initial velocities, and the parameters $\delta=0$, 
$\gamma=1$ and $\zeta=-1.7774$. The blue shaded region represents the first stage of inflation, and the green shaded region represents the second stage of inflation. (b) The trajectory of the solution. (c) The corresponding Hubble function. (d) The e-folds.  (e) The SR parameters $\epsilon$ (red) and 
$\eta$ (blue).}
\label{fsV_sol_gamma}
\end{figure}

The power spectrum of curvature perturbations is numerically computed at fixed $\Delta N_2$ by using the standard transport method~\cite{Mulryne:2009kh,Mulryne:2010rp} with the Mathematica package~\cite{Dias:2015rca}, around the pivot scale $k_*$ that leaves the horizon at the end of the first stage (we call this scale $k_{\Delta N_2}$). The inflaton mass is chosen to be $0.5\times 10^{-5}M_{\rm Pl}$ by requiring $P_\zeta\approx 2\times 10^{-9}$ for the mode $k$ that exits the horizon 60 e-folds before the end of inflation (we call it $k_{60}$). The results for various values of $\gamma$  are shown in Fig.~\ref{Pk_int_gamma}. The values of the parameters are collected in Table \ref{tab_gamma}, where $\zeta$ is tuned to satisfy $\Delta N_2\approx 10$. 

\begin{figure}
\centering
  \includegraphics[width=.55\linewidth]{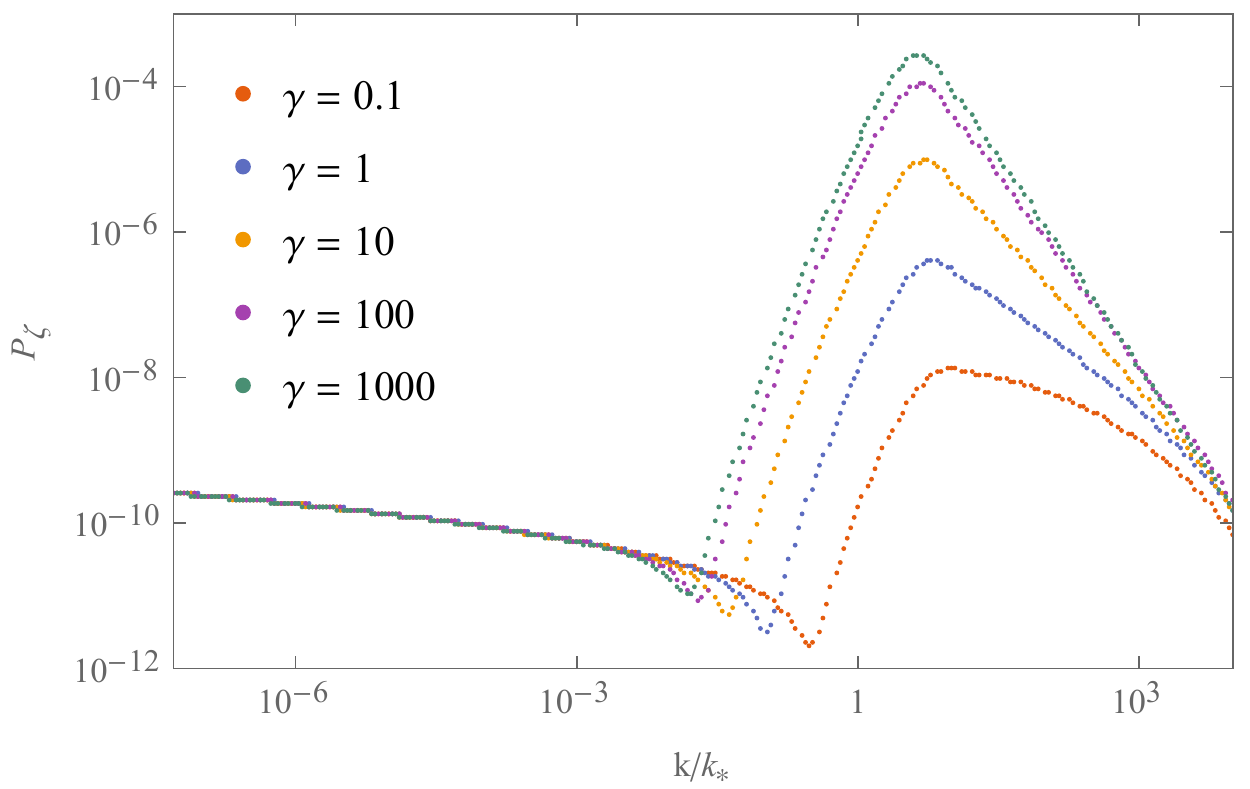}
\captionsetup{width=.9\linewidth}
\caption{The power spectrum $P_{\zeta}$ near the pivot scale $k_*=k_{\Delta N_2}$ at $\Delta N_2=10$ for some values of $\gamma$.}
\label{Pk_int_gamma}
\end{figure}

\begin{table}[ht]
\centering
\begin{tabular}{l r r r r r}
\toprule
$\gamma$ & $0.1$ & $1$ & $10$ & $100$ & $1000$\\
$\zeta$ & $-0.31165$ & $-1.7774$ & $-8.91495$ & $-42.7976$ & $-201.722$\\\bottomrule
\hline
\end{tabular}
\captionsetup{width=.9\linewidth}
\caption{The parameters leading to the power spectrum in Fig.~\ref{Pk_int_gamma} with $\Delta N_2\approx 10$.}
\label{tab_gamma}
\end{table}

The enhancement of primordial curvature perturbations needed for PBH formation is usually assumed to be  
$\frac{P_{\zeta,{\rm max.}}}{P_{\zeta,{\rm min.}}}\equiv P_{\rm enh.}\sim 10^{7}$, where the subscripts refer to the values of $P_{\zeta}$ at the peak and the base of the peak, respectively, in comparison to the CMB scale.  Given a {\it broad} peak, the enhancement may be less by one order of the magnitude \cite{Germani:2018jgr}. According to Fig.~\ref{Pk_int_gamma}, when choosing $\Delta N_2=10$, the enhancement $P_{\rm enh.}\gtrsim 10^6$ is achieved for $\gamma\gtrsim 10$. Actually, it ranges from $10^6$ at  $\gamma=10~$ to $10^7$ 
at $\gamma=1000$. 

In Table~\ref{tab_gamma_dn} the values of $n_s$ and $r_{\rm max.}$ are collected at the CMB scales for the values of $\Delta N_2=10,17,20,23$, universally across the considered values of 
$\gamma=0.1,1,10,100,1000$. The tensor-to-scalar ratio $r$ is well within the observational limits in all those cases, but the scalar tilt $n_s$ is outside the $1\sigma$ limit when $\Delta N_2=20$, and is marginally outside the $3\sigma$ limit when $\Delta N_2=23$.

\begin{table}[ht]
\centering
\begin{tabular}{l r r r r}
\toprule
$\Delta N_2$ & $10$ & $17$ & $20$ & $23$ \\
\hline
$n_s$ & $0.955$ & $0.946$ & $0.942$ & $0.936$ \\
$r_{\rm max}$ & $0.006$ & $0.008$ & $0.009$ & $0.011$ \\\bottomrule
\hline
\end{tabular}
\captionsetup{width=.9\linewidth}
\caption{The approximate values of $n_s$ and $r$ for some choices of $\Delta N_2$.}
\label{tab_gamma_dn}
\end{table}

The mass of PBH created as a result of the primordial power spectrum enhancement can be estimated  from the peak data as follows~\cite{Pi:2017gih}:
\begin{eqnarray}
    M_{\rm PBH}\simeq \fracmm{M^2_{\rm Pl}}{H(t_{\rm peak})}\exp\left[2(N_{\rm end}-N_{\rm peak})+\int^{t_{60}}_{t_{\rm peak}}\epsilon(t)H(t)dt\right]~,\label{MPBH}
\end{eqnarray}
where $t_{\rm peak}$ is the time when the wavenumber corresponding to the power spectrum peak ($k_{\rm peak}$) exits the horizon, and $t_{60}$ is the time when $k_{60}$ exits the horizon. 

The values of $M_{\rm PBH}$ for some values of $\Delta N_2$ from Eq.~\eqref{MPBH} are shown in Table~\ref{tab_MPBH_gamma}  together with the corresponding values of the spectral index. Those estimates are universal across the values of $\gamma=0.1,1,10,100,1000$. On the one hand, PBH with masses smaller than $\sim 10^{16}{\rm g}$ should have already evaporated until now via Hawking radiation. Hence, we require $\Delta N_2 \geq 20$. On the other hand, the lower $3\sigma$ limit on the spectral index, $n_s\approx 0.946$, requires $\Delta N_2<23$. Hence, PBH masses are restricted in our models by ${\cal O}(10^{16}{\rm g})<M_{\rm PBH}<{\cal O}(10^{19}{\rm g})$ before imposing current observational constraints on them.

\begin{table}[ht]
\centering
\begin{tabular}{l r r r r}
\toprule
$\Delta N_2$ & $10$ & $17$ & $20$ & $23$ \\
\hline
$M_{\rm PBH}$, g & $10^{9}$ & $10^{15}$ & $10^{17}$ & $10^{20}$ \\
$n_s$ & $0.955$ & $0.946$ & $0.942$ & $0.936$ \\\bottomrule
\hline
\end{tabular}
\captionsetup{width=.9\linewidth}
\caption{The PBH masses from Eq.~\eqref{MPBH} in the $\gamma$ model and the values of the scalar spectral index $n_s$. In the Solar mass units, $1~{\rm g}\approx 5.03\times 10^{-34}~M_\odot$.}
\label{tab_MPBH_gamma}
\end{table}

The PBH density fraction in DM can be estimated by using the standard (Press-Schechter) formalism \cite{Press:1973iz}. The useful formulae include the PBH mass $\tilde{M}_{\rm PBH}(k)$, the production rate $\beta_f(k)$, and the density contrast $\sigma(k)$ coarse-grained over $k$ as follows (see e.g., Refs.~\cite{Inomata:2017okj,Inomata:2017vxo} and the references therein):
\begin{gather}
    \tilde{M}_{\rm PBH}\simeq 10^{20}\left(\fracmm{7\times 10^{12}}{k~{\rm Mpc}}\right)^2{\rm g}~, \quad \beta_f(k)\simeq\fracmm{\sigma(k)}{\sqrt{2\pi}\delta_c}
    e^{-\fracmm{\delta^2_c}{2\sigma^2(k)}}~,\\ \nonumber
    \sigma^2(k)=\fracmm{16}{81}\int\fracmm{dq}{q}\left(\fracmm{q}{k}\right)^4e^{-q^2/k^2}P_\zeta(q)~.\label{PBH_productionE}
\end{gather}

We have chosen the Gaussian window function for the density contrast, and have introduced $\delta_c$ as a constant representing the density threshold for PBH formation. It is often assumed that $\delta_c\approx 1/3$ \cite{Carr:1975qj}, though it may be different. The PBH-to-DM density fraction is estimated as \cite{Inomata:2017okj,Inomata:2017vxo}
\begin{eqnarray}
    \fracmm{\Omega_{\rm PBH}(k)}{\Omega_{\rm DM}}\equiv f(k)\simeq\fracmm{1.4\times 10^{24}\beta_f(k)}{\sqrt{\tilde{M}_{\rm PBH}(k){\rm g}^{-1}}}~~.\label{f_PBH}
\end{eqnarray}

In order to numerically evaluate the function \eqref{f_PBH}, we have normalized the values of $k$ in terms of the observable scales today, and the scale $k_{60}$ was chosen to represent the largest currently observable scale around $10^{-4}~{\rm Mpc}^{-1}$. Then our numerical calculation reveals that the $\delta_c$ parameter should be a bit smaller than $1/3$ (actually, close to $\delta_c=0.275$ in the case under consideration).

The $\gamma$-models of inflation and PBH formation studied above have rather low values of the CMB spectral index  (with $n_s\approx 0.942$ as the best fit) which are in tension ($3\s$) with the precision measurements of Planck mission \cite{Planck:2018jri} in Eq.~(\ref{PlanckCMB}). The better values are found in Subsection 3.3.

\subsection{The $\delta$ models}

The $\delta$ models, defined by $\gamma=0$ and $\delta\neq 0$ in Eqs.~(\ref{N_choice2}) and (\ref{N_F_choice2}), break the R-symmetry and the reflection symmetry $\sigma\rightarrow -\sigma$ of the potential, see Fig.~\ref{V_3d_delta}. For any non-zero $\delta$, there is the value of $\zeta$ leading to an inflection point.

\begin{figure}
\centering
\begin{subfigure}{.4\textwidth}
  \centering
  \includegraphics[width=1\linewidth]{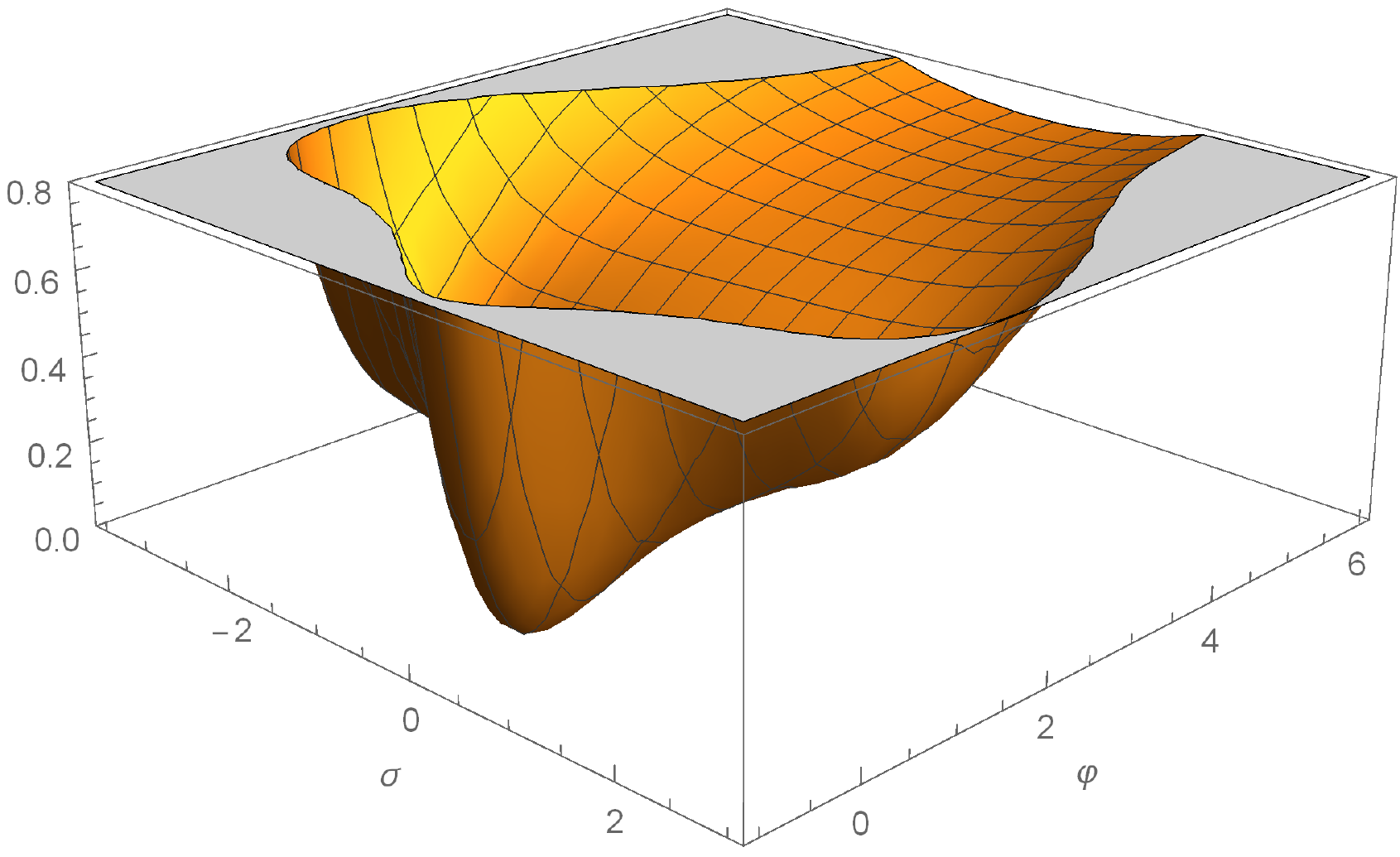}
  \label{V_3d1_delta}
\end{subfigure}
\begin{subfigure}{.49\textwidth}
  \centering
  \includegraphics[width=.85\linewidth]{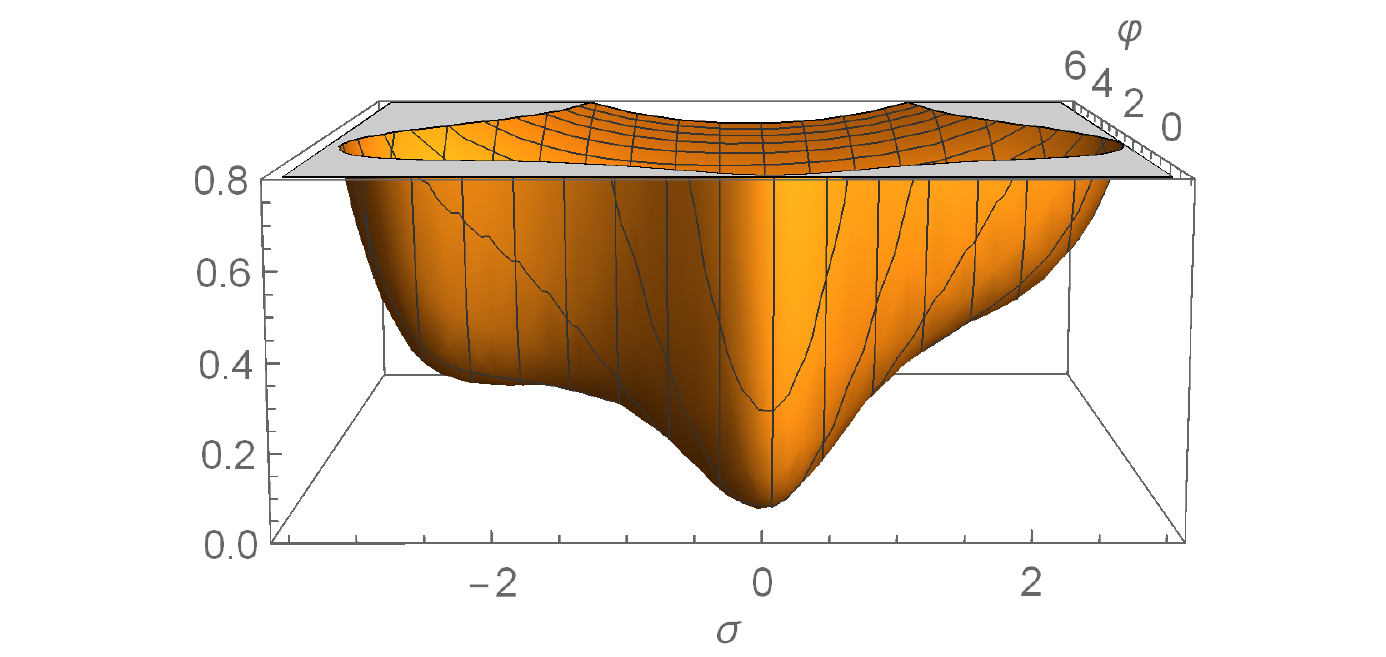}
  \label{V_3d2_delta}
\end{subfigure}
\captionsetup{width=.9\linewidth}
\caption{The scalar potential for $\gamma=0$, $\delta=0.1$ and $\zeta=0.033407$.}
\label{V_3d_delta}
\end{figure}

In contrast to the $\gamma$ models, there is a single valley for large positive $\varphi$ and $\sigma=0$. When approaching $\varphi=0$,  the inflationary trajectory passes the (near-)inflection point and then falls to the Minkowski minimum at $\varphi=\sigma=0$.

Let us take the parameter values $\delta=0.1$ and $\zeta=0.033407$, where $\zeta$ is chosen to get $\Delta N_2=10$. A numerical solution to the field equations yields the time dependence of $\varphi$, $\sigma$, $\tilde{H}$, $N$, $\epsilon$ and $\eta$ shown in Fig.~\ref{fsV_sol_delta}. The near-inflection point divides inflation into two stages with $\Delta N_1=50$ and $\Delta N_2=10$, respectively. The initial velocities are set to zero, with $\varphi(0)=6$ and $\sigma(0)=0.05$. Similarly to the $\gamma$ models, the inflationary trajectory is (locally) stable against variations of the initial conditions.

\begin{figure}
\centering
\begin{subfigure}{.49\textwidth}
  \centering
  \includegraphics[width=0.85\linewidth]{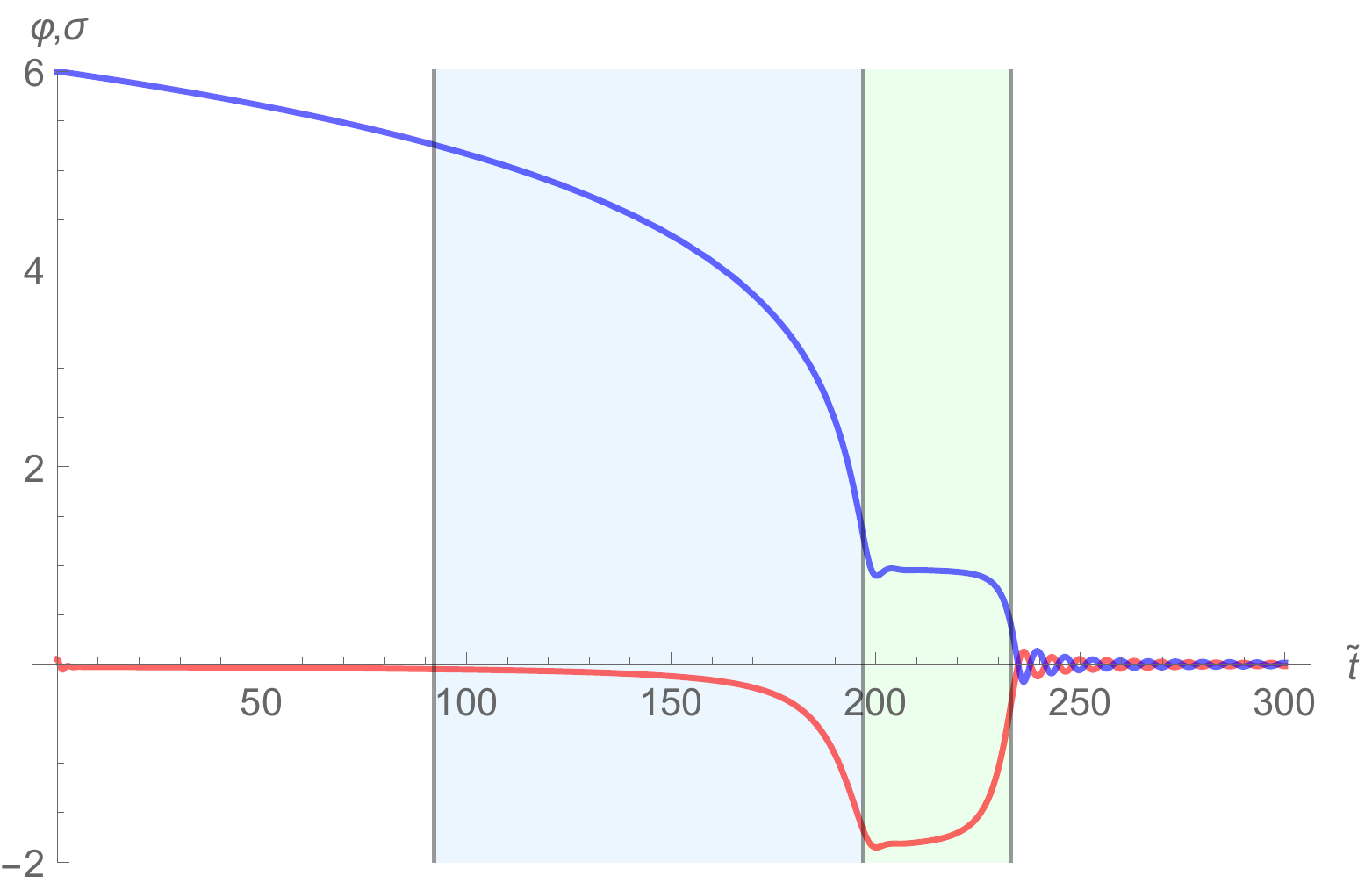}
  \caption{}
  \label{fs_sol_delta}
\end{subfigure}
\begin{subfigure}{.49\textwidth}
  \centering
  \includegraphics[width=.75\linewidth]{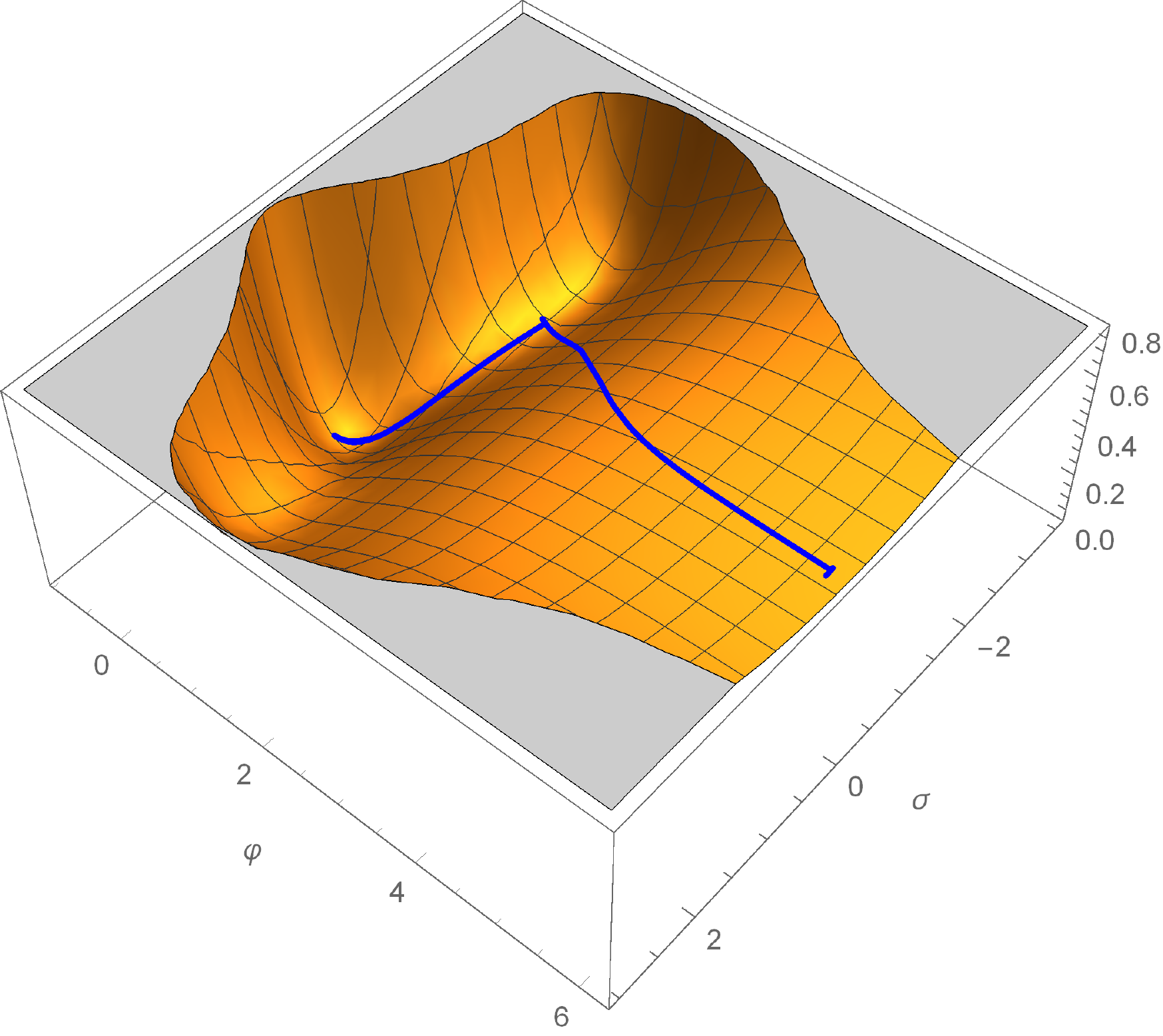}
  \caption{}
  \label{V_sol_delta}
\end{subfigure}
\begin{subfigure}{.32\textwidth}
  \centering
  \includegraphics[width=.9\linewidth]{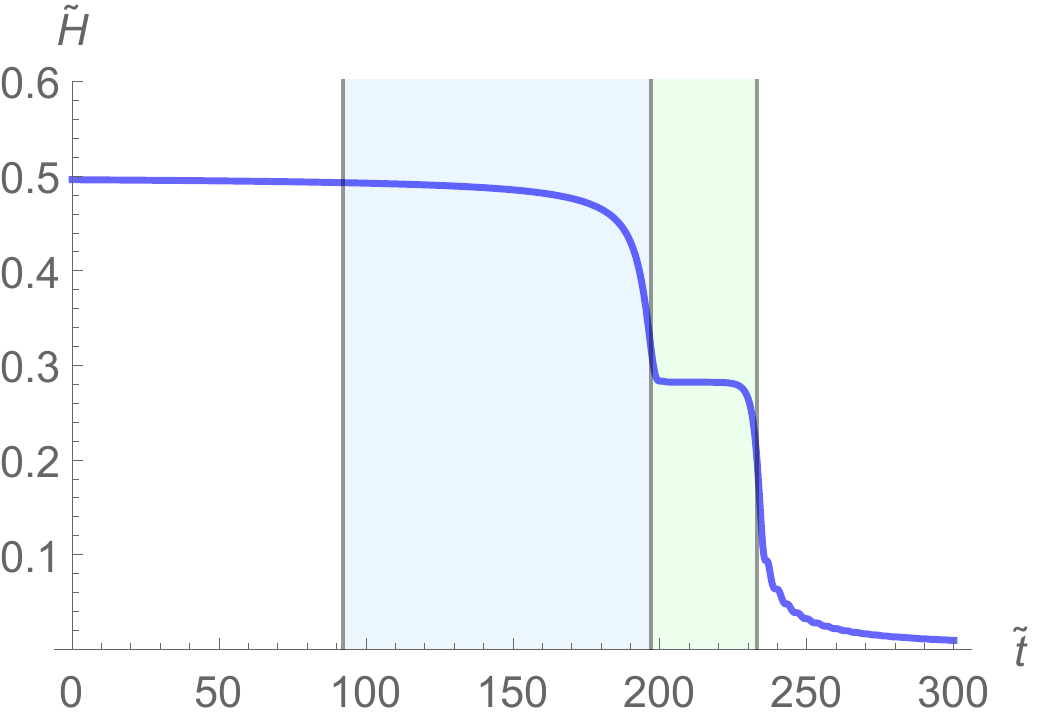}
  \caption{}
  \label{H_sol_delta}
\end{subfigure}
\begin{subfigure}{.32\textwidth}
  \centering
  \includegraphics[width=.9\linewidth]{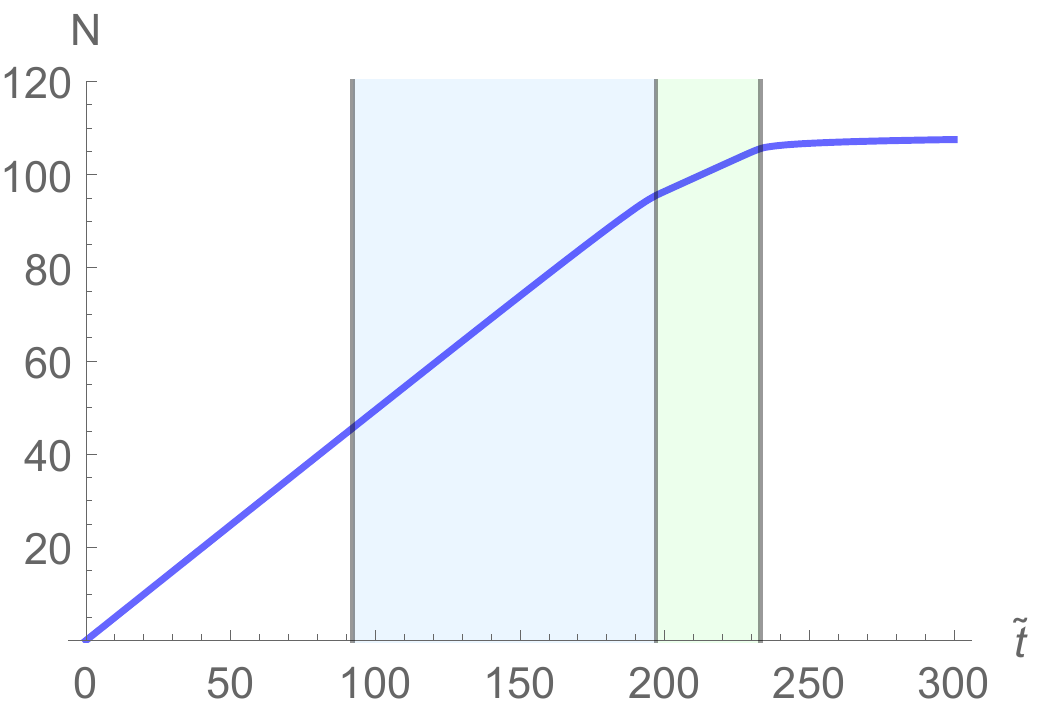}
  \caption{}
  \label{N_sol_delta}
\end{subfigure}
\begin{subfigure}{.32\textwidth}
\centering
\includegraphics[width=.95\linewidth]{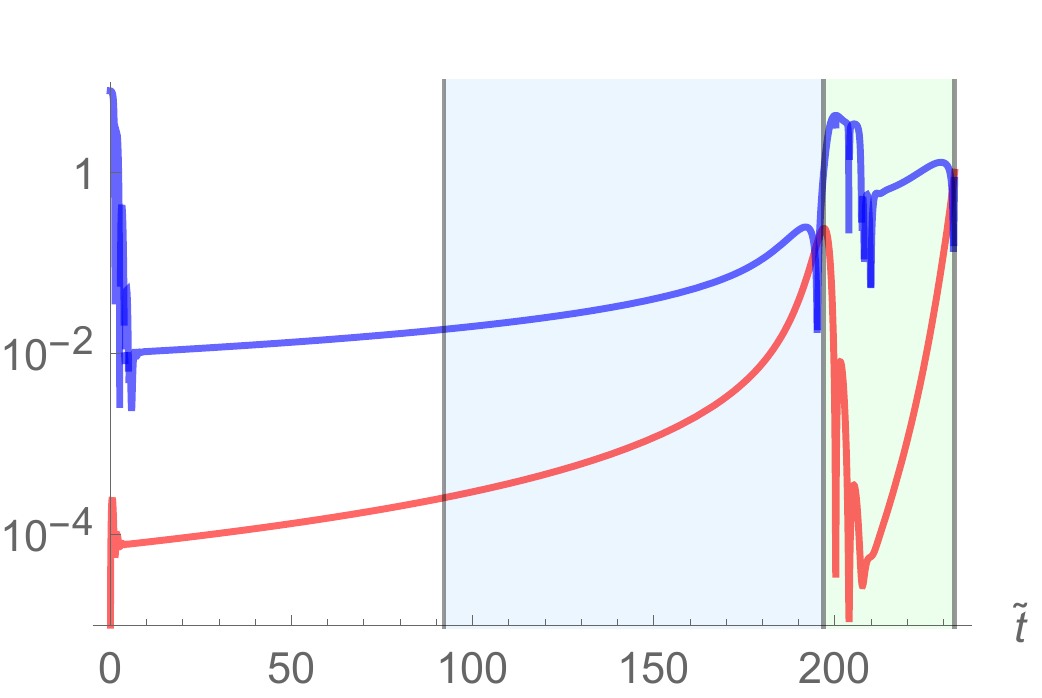}
\captionsetup{width=.9\linewidth}
\caption{}
\label{en_sol_delta}
\end{subfigure}
\captionsetup{width=.9\linewidth}
\caption{(a) The solution to the field equations \eqref{KG1_gamma} and \eqref{KG2_gamma} with the initial conditions $\varphi(0)=6$ and $ \sigma(0)=0.05$, the vanishing  initial velocities, and the parameter choice $\gamma=0$, $\delta=0.1$ and  $\zeta=0.033407$. (b) The trajectory of the solution ($\varphi$ -- blue, $\sigma$ -- red). (c) The Hubble function. (d) The e-folds. (e) The slow-roll parameters $\epsilon$ (red) and $\eta$ (blue).}
\label{fsV_sol_delta}
\end{figure}

The power spectrum enhancement also takes place, either with a {\it smooth} peak for $\delta\sim 0.1$, or with a {\it sharp} peak for $\delta\sim 0.6$, in the scalar power spectrum. When requiring the PBH density fraction \eqref{f_PBH} to be close to one and the corresponding density threshold not to deviate far away from the region $1/3\leq\delta_c\leq 2/3$, one finds that the values $\delta=0.094$ and $\delta=0.58$,
respectively, are suitable for efficient generation of PBH with their masses around $10^{19}{\rm g}$, while avoiding their overproduction ($f\lesssim 1$).

The results about the $\delta$-models are summarized in Table \ref{tab_MPBH_delta}.  To get $M_{\rm PBH}\sim 10^{19}{\rm g}$ one should either set $\delta=0.094$ and $\Delta N_2=20$ (in this case $P_{\rm enh}\approx 4.5\times 10^7$), or $\delta=0.58$ and $\Delta N_2=23$ (in this case $P_{\rm enh}\approx 2.7\times 10^8$). In the former case, $n_s$ is within $2\sigma$ CL, whereas in the latter case, $n_s$ is within $3\sigma$ but outside $2\sigma$ CL. Therefore, those $\delta$-models of inflation and PBH formation are better than the $\gamma$ models, as regards matching the predicted values of the CMB scalar perturbations index $n_s$ with Planck measurements \cite{Planck:2018jri}. 

\begin{table}[ht]
\centering
\begin{tabular}{l r r r r r r r r}
\toprule
& \multicolumn{4}{c}{$\delta=0.09$} & \multicolumn{4}{c}{$\delta=0.61$}\\
\cmidrule(r){2-5} \cmidrule(l){6-9}
$\Delta N_2$ & $10$ & $17$ & $20$ & $23$ & $10$ & $17$ & $20$ & $23$ \\
\hline
$M_{\rm PBH}$, g & $10^{9}$ & $10^{15}$ & $10^{18}$ & $10^{20}$ & $10^{9}$ & $10^{15}$ & $10^{18}$ & $10^{20}$ \\
$n_s$ & $0.9566$ & $0.9486$ & $0.9443$ & $0.9390$ & $0.9581$ & $0.9504$ & $0.9461$ & $0.9409$\\
$r_{\rm max}$ & $0.005$ & $0.007$ & $0.008$ & $0.010$ & $0.004$ & $0.006$ & $0.007$ & $0.008$ \\\bottomrule
\hline
\end{tabular}
\captionsetup{width=.9\linewidth}
\caption{The PBH masses from Eq.~\eqref{MPBH} for $\delta=0.09$ and $\delta=0.61$, with the corresponding values of $n_s$ and $r_{\rm max}$.}
\label{tab_MPBH_delta}
\end{table}

\subsection{Comparison with observational constraints on PBH and DM}

The specific models studied in the preceding Subsections can be further improved and compared to the current observational  constraints on PBH and DM, after fine tuning their parameters to the best fit within the modified Starobinsky-type supergravity \cite{Aldabergenov:2020bpt,Aldabergenov:2020yok}. 

For those purposes we pick up a $\gamma$-model (Case I) and two $\delta$-models (Cases II and III) with the different shapes of the power spectrum.  The choice of two $\delta$-models is motivated by the existence of two suitable parameter regions, where $\delta\simeq 0.1$ and $\delta\simeq 0.6$ yield truly different shapes of the power spectrum (broad and narrow peaks), respectively. The parameters of those three models are given in Table \ref{Tab_eg}, and the corresponding (numerically computed) power spectra $P_\zeta$ and PBH density fractions $f(M)$ are shown in Fig.~\ref{Fig_Pf} with the normalization of the wavenumber $k_{\rm exit}=0.05~{\rm Mpc}^{-1}$, where $k_{\rm exit}$ is the scale that leaves the horizon around $54$ e-folds. The parameter $\zeta$ is fixed by choosing the value of $\Delta N_2$ at given $\gamma$ and $\delta$. In the cases I, II and III, the parameter $\zeta$ is given by $-2.374$, $0.032$, and $0.102$, respectively.

\begin{table}[ht]
\centering
\begin{tabular}{l r r r r r r}
\toprule
& $\gamma$ & $\delta$ & $\Delta N_2$ & $\delta_c$ & $n_s$ & $r$ \\
\hline
Case I & $1.5$ & $0$ & $20$ & $0.4$ & $0.942$ & $0.009$\\
Case II & $0$ & $0.09$ & $19$ & $0.47$ & $0.946$ & $0.008$\\
Case III & $0$ & $0.61$ & $20$ & $0.4$ & $0.946$ & $0.007$\\
\hline
\end{tabular}
\captionsetup{width=.9\linewidth}
\caption{The parameters corresponding to the PBH density fraction in Fig.~\ref{Fig_Pf}. The $n_s$ and $r$ are computed with the total $\Delta N=54$ e-folds before the end of inflation.}
\label{Tab_eg}
\end{table}

\begin{figure}[htb]
\begin{subfigure}{.49\textwidth}
 \centering
 \includegraphics[width=1\linewidth]{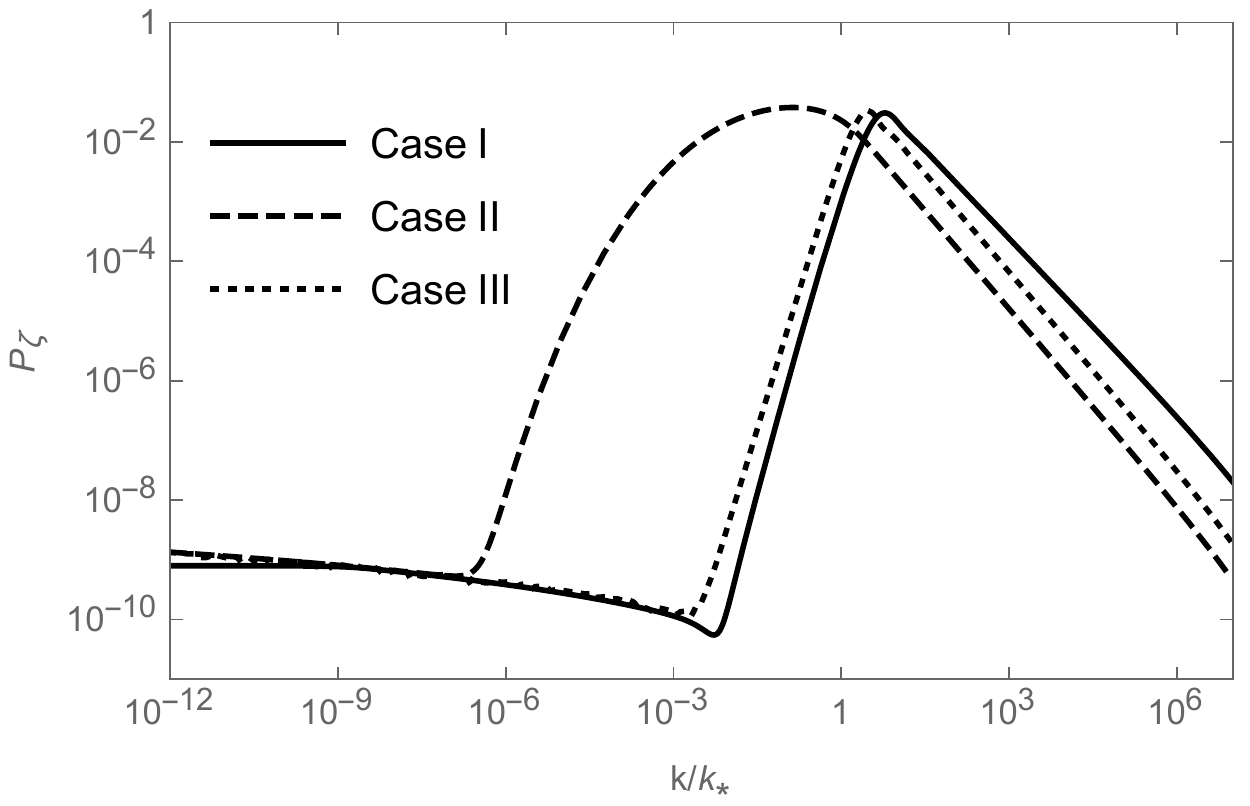}
\end{subfigure}
\begin{subfigure}{.49\textwidth}
  \centering
  \includegraphics[width=1\linewidth]{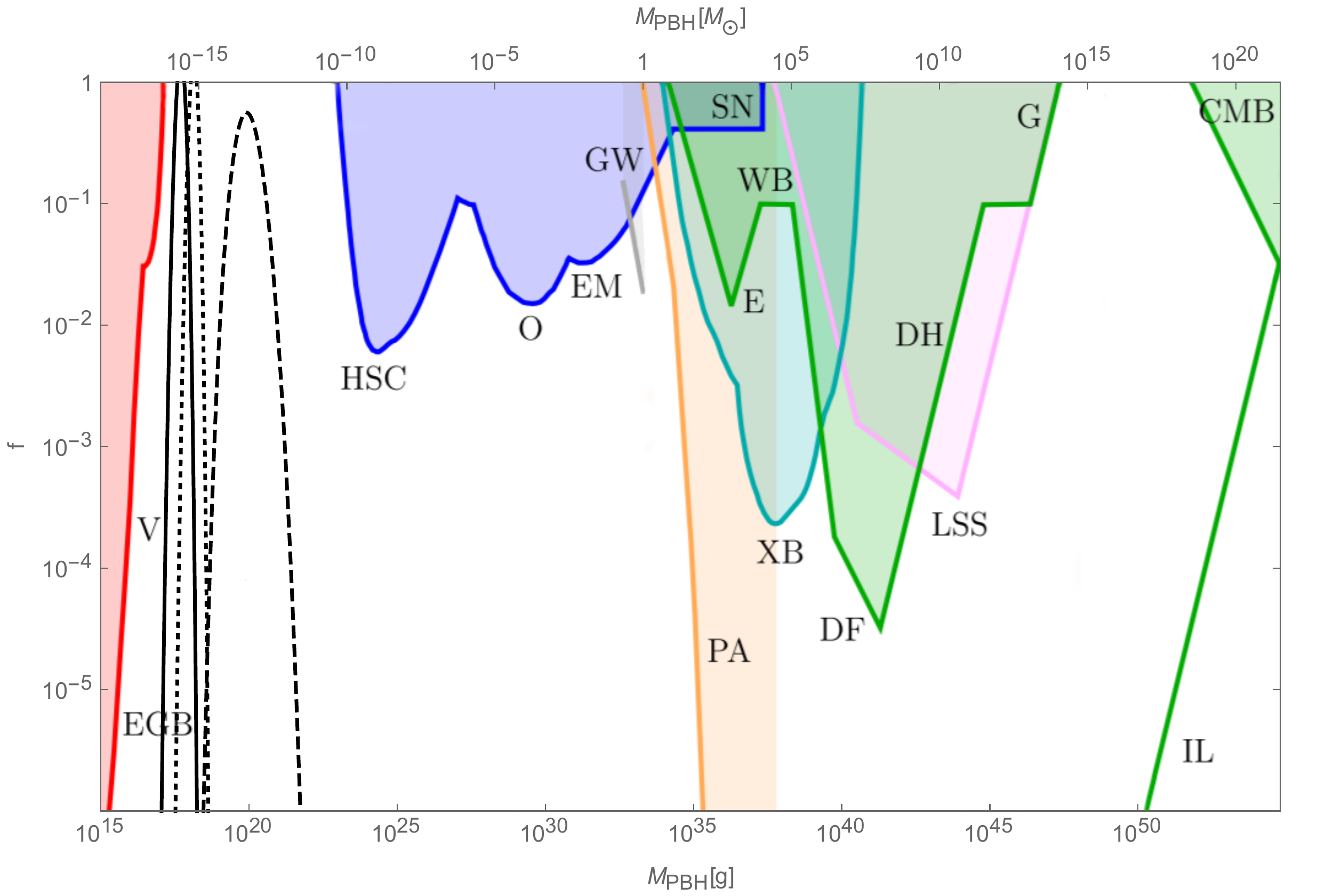}
\end{subfigure}
\captionsetup{width=.9\linewidth}
\caption{The power spectrum in the three examples of Table \ref{Tab_eg}, see the left side. The $k_*$ represents the end of SR and the beginning of USR. The corresponding PBH density fractions are given on the right side. The observational constraints on PBH in the background (on the right side) are taken from Refs.~\cite{Carr:2020gox,Carr:2020xqk}. In both plots, the case I is denoted by the solid line, the case II by  the dashed line, and the case III by the dotted line. The whole DM composed from PBH is possible in the cases I and III.}
\label{Fig_Pf}
\end{figure}

According to Table \ref{Tab_eg}, the spectral tilt $n_s$ in the case I is in tension by $3\sigma$ against the CMB data \cite{Planck:2018jri}, whereas in the cases II and III the value of $n_s$ is within the current $3\sigma$ constraints.  Therefore, the $\delta$-models are more suitable to accommodate the observed values of $n_s$, though with the PBH fraction less than the whole DM. In particular, the PBH fraction in the case II peaks at the center of the allowed window,  while it is still possible to move the peak further to the left, thus lowering the PBH masses. It is also worth noticing that we have used the value of the critical $\d_c$ parameter beyond $1/3$, see Table 5.

  \section{Scanning the parameter space}

 In this Section we study the parameter space of the models defined by Eqs.~(\ref{N_choice2}) and (\ref{N_F_choice2}), with both $\gamma$ and $\delta$ to be non-vanishing. By searching for the proper values of those parameters,  we investigate whether it is possible to move the value of the cosmological observable $n_s$ into the most favorable region  (within $1\s$), while keeping the efficient PBH production.  We also study some new models by changing the second (quadratic) term in our Ansatz (\ref{N_F_choice2}) to a higher power.

 \subsection{The $(\gamma,\delta)$ models}
 
 In the $(\gamma,\delta)$ models, the parameter values should be restricted in order to get a sufficient length of the second stage of inflation, required for efficient PBH production (as DM). In addition, with the parameter values similar to those  in the previous Section, the USR phase between the first and the second plateaus of the inflationary trajectory requires extreme fine-tuning that should be avoided.  As a fully analytic approach is impossible, we scan the parameter space {\it randomly} by doing numerical calculations in our search for better agreement with the Planck measurements of $n_s$.
 
 \begin{figure}
  \centering
    \includegraphics[width=.55\linewidth]{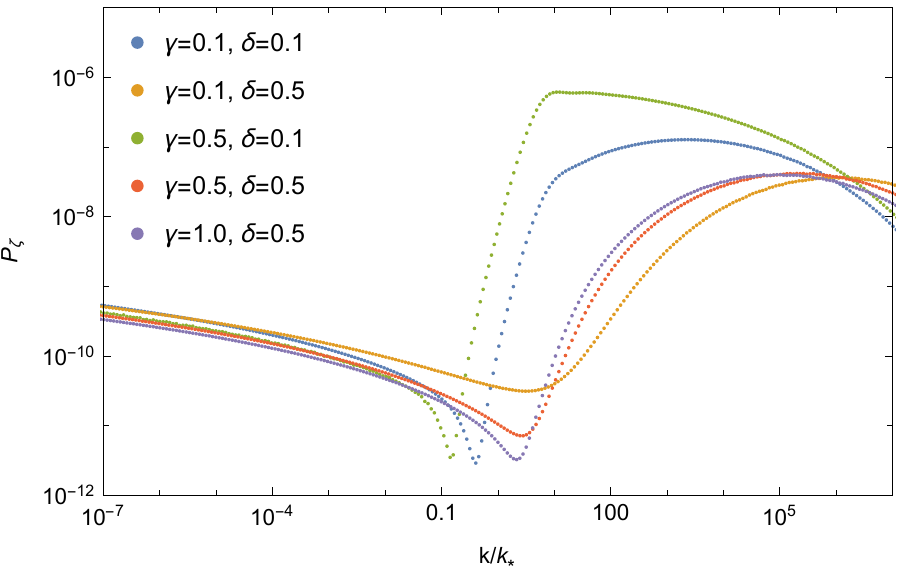}
  \captionsetup{width=.9\linewidth}
  \caption{The power spectrum $P_{\zeta}$ around the pivot scale $k_*$ at $\Delta N_2\approx 20$ for the selected values of $\gamma$ and $\delta$.}
  \label{fig_gamma_delta_p}
\end{figure}

Figure~\ref{fig_gamma_delta_p} shows the power spectrum $P_{\zeta}$ around the pivot scale $k_{*}$ for some non-vanishing values of $\gamma$ and $\delta$ given in Table \ref{tab_gamma_delta}, which were used (separately)  in the preceding Section.

In Table \ref{tab_gamma_delta} we also collect the values of the parameters $\gamma$, $\delta$ and $\zeta$, where  $\zeta$ has been tuned  to get $\Delta N_2 \approx 20$, together with the values of the power spectrum enhancement $P_{\rm enh.}$  (relative to CMB)  for them, as our first (naive) trials.

Like the special cases of the preceding Section, going well below $20$ for $\Delta N_2$ leads to an increase of $n_s$
accompanied by a drastic decrease in the PBH masses (significantly below the Hawking radiation limit) and a much lower enhancement of the scalar power spectrum, see Fig.~\ref{pow_n2deltaN}.  We assume $\Delta N_2 \approx 20$  in what follows.

 As can be seen from Table~\ref{tab_gamma_delta}, the adjustment of the value of $\zeta$ to get $\Delta N_2 \approx 20$ is generically possible in a large region in the parameter space $(\gamma,\delta)$, while  the required value of $\zeta$ mainly depends on the value of $\gamma$. 

 \begin{table}[ht]
  \centering
  \begin{tabular}{l r r r r r}
  \toprule
  $\gamma$ & $0.1$ & $0.1$ & $0.5$ & $0.5$ & $1.0$ \\
  $\delta$ & $0.1$ & $0.5$ & $0.1$ & $0.5$ & $0.5$ \\
  $\zeta$ & $-0.3335$ & $-0.3934$ & $-1.1137$ & $-1.2280$ & $-1.9826$ \\
  $\Delta N_2$ & $20.23$ & $20.11$ & $20.35$ & $20.39$ & $20.06$ \\
  $P_{\rm enh.}$ & $4.4\times 10^4$ & $1.2\times 10^3$ & $1.7\times 10^5$ & $5.8\times 10^3$ & $1.2\times 10^4$ \\
  \bottomrule
  \end{tabular}
  \captionsetup{width=.9\linewidth}
  \caption{The selected values of the parameters $\gamma$, $\delta$ and $\zeta$ leading to $\Delta N_2 \approx 20$, and the power spectrum enhancement $P_{\rm enh.}$ (relative to CMB), respectively.}
  \label{tab_gamma_delta}
  \end{table}

As is clear from Table~\ref{tab_gamma_delta} and Fig.~\ref{fig_gamma_delta_p}, we observe the sharp decrease in the scalar power spectrum enhancement, well below $10^6$. Therefore, stricter restrictions on the values of the parameters $\gamma$ and $\delta$ are necessary, while their existence for efficient PBH production (as DM) seems to be non-trivial. It signals a qualitative departure from the two special cases considered in the preceding Section.

\begin{figure}[t]
\centering
\includegraphics[width=.55\linewidth]{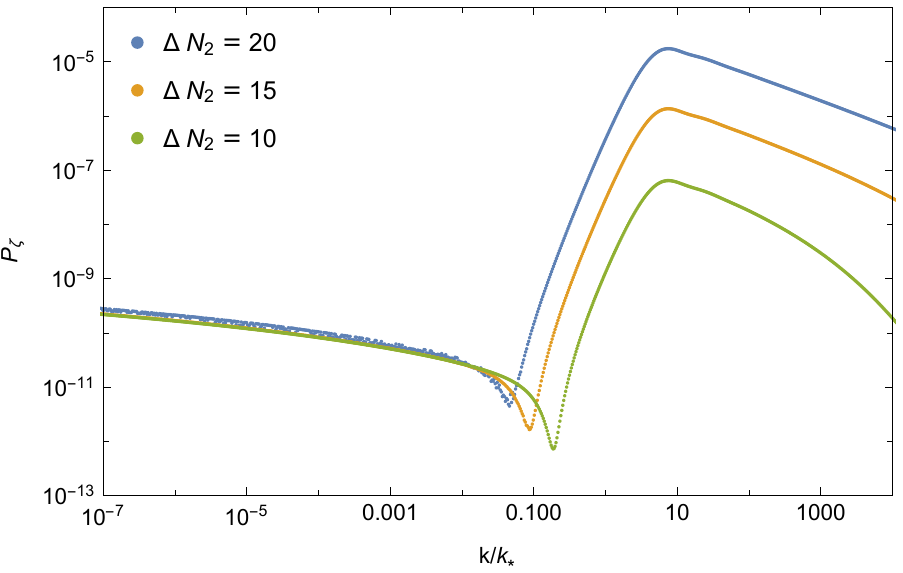}
\captionsetup{width=.9\linewidth}
\caption{The power spectrum $P_{\zeta}$ near the pivot scale $k_*=k_{\Delta N_2}$
for some values of $\Delta N_2$ less or equal $20$.}
\label{pow_n2deltaN}
\end{figure}

Afer studying the power spectra around $\gamma,\delta \approx 1.0$ with $\Delta N_2 \approx 20$, we find that their enhancement (peak) becomes larger when $\gamma$ increases and $\delta$ decreases, though only slightly. The sufficient enhancement of primordial curvature perturbations, needed for the efficient PBH formation, is given by $P_{enh.}\sim 10^6$ at least \cite{Germani:2018jgr}. It follows from our numerical calculation and animation that the value of $\gamma$ in the models with $\delta\geq 0.1$ should be as large as $10$ at least. Accordingly, we are led to the representative values of $\gamma$ and $\delta$ as $\gamma=10$, $\delta=0.2$ and adjust $\zeta=-9.15915$. 
The shape of the (canonical) scalar potential in that $(\gamma,\delta)$ model is shown in  Fig.~\ref{fig_gamma_delta_V}. As may have been expected, it combines the shapes of the potentials observed in 
the $\gamma$- and $\delta$-models, see Figs.~\ref{V_3d_gamma} and \ref{V_3d_delta}, respectively. In
particular, the potential in Fig.~\ref{fig_gamma_delta_V} has the same pattern of the reflection symmetry breaking seen in the $\delta$ models.

The evolution of the scalars is given by a solution to the equations of motion in the FLRW spacetime, and it is displayed in Fig.~\ref{fig_gamma_delta}. The evolution of Starobinsky's scalaron $\varphi$ and the SR parameter $\epsilon$ are shown by the red lines, whereas the evolution of the field $\sigma$, the SR parameter $\eta$ and the Hubble function $H$ are shown by the blue lines. We use the initial conditions $\varphi(0)=6$ and $\sigma(0)=0.1$, with the vanishing velocities. Should the initial condition for the $\sigma$-field be with the opposite sign, say, $\delta(0)=-0.1$, there will be no second plateau, no second stage of inflation and, hence, no PBH production.

\begin{figure}
  \centering

  \begin{subfigure}{.49\textwidth}
    \centering
    \includegraphics[width=.85\linewidth]{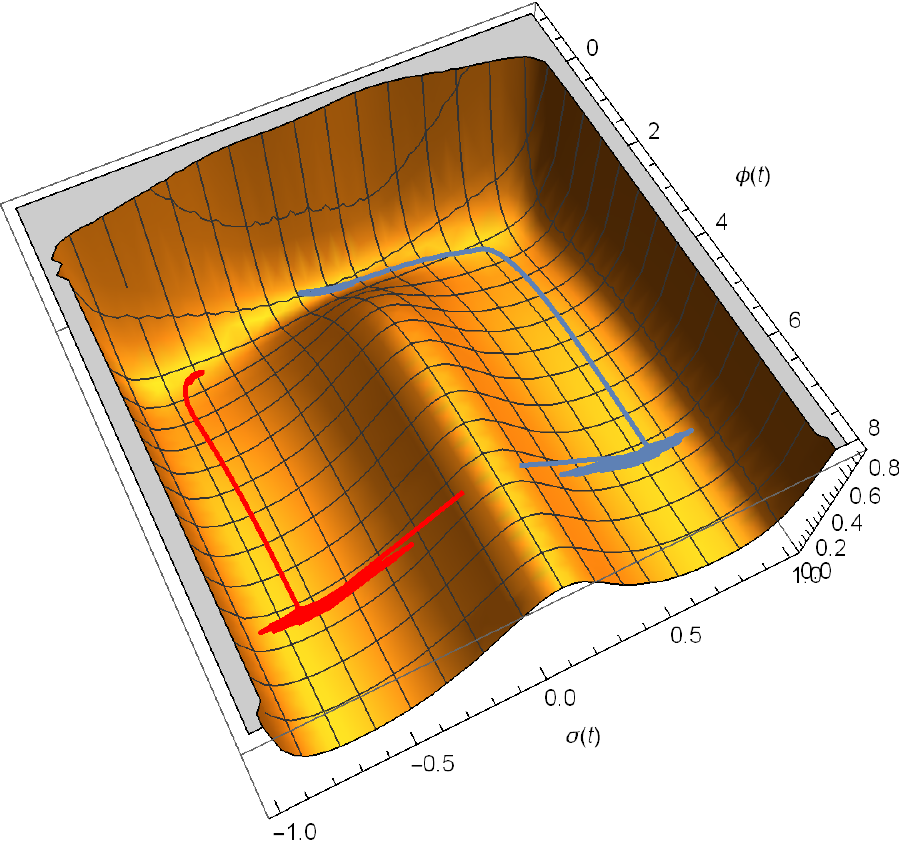}
    \caption{}
    \label{fig_gamma_delta_V}
  \end{subfigure}
  \begin{subfigure}{.49\textwidth}
    \centering
    \includegraphics[width=.9\linewidth]{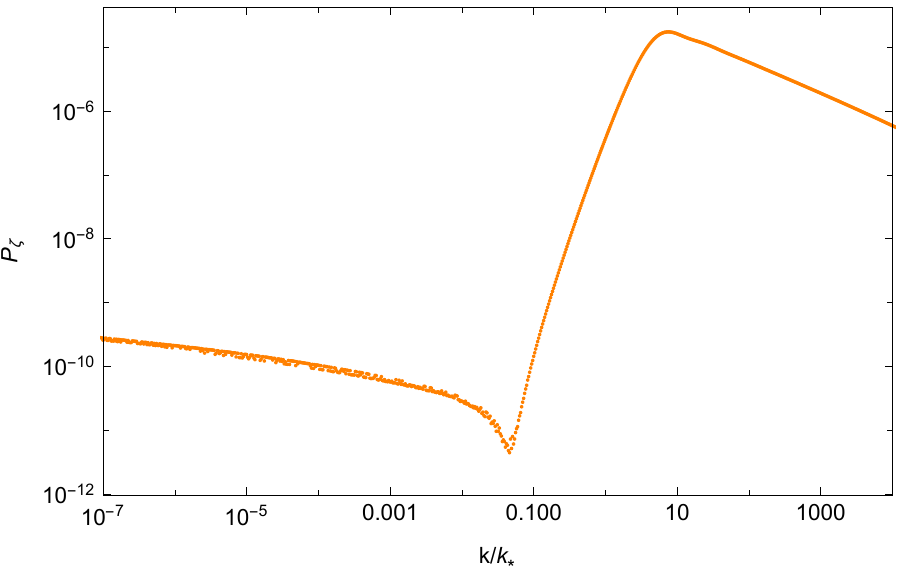}
    \caption{}
    \label{fig_gamma_delta_N}
  \end{subfigure}
  \begin{subfigure}{.32\textwidth}
    \centering
    \includegraphics[width=1\linewidth]{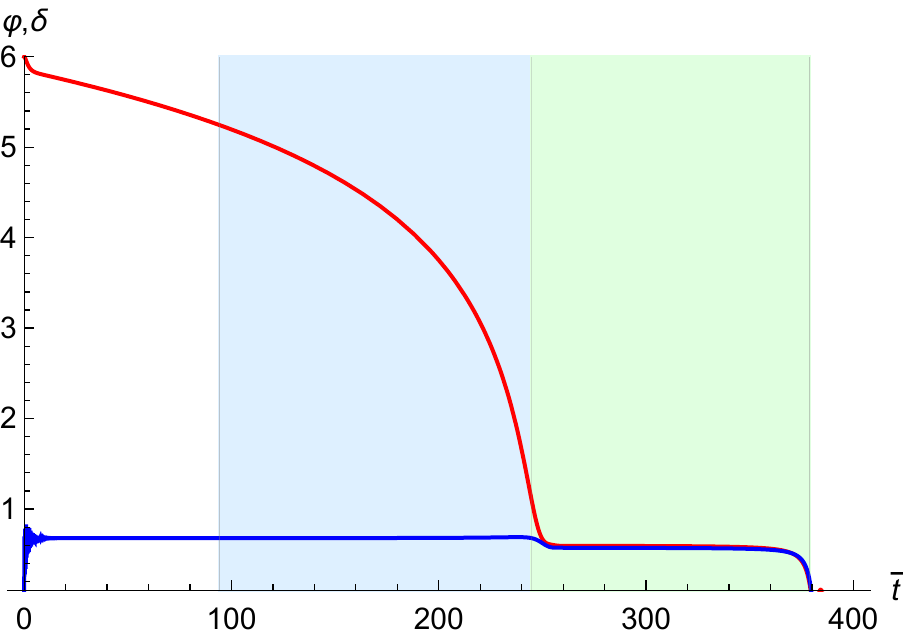}
    \caption{}
    \label{fig_gamma_delta_sol}
  \end{subfigure}
  \begin{subfigure}{.32\textwidth}
    \centering
    \includegraphics[width=1\linewidth]{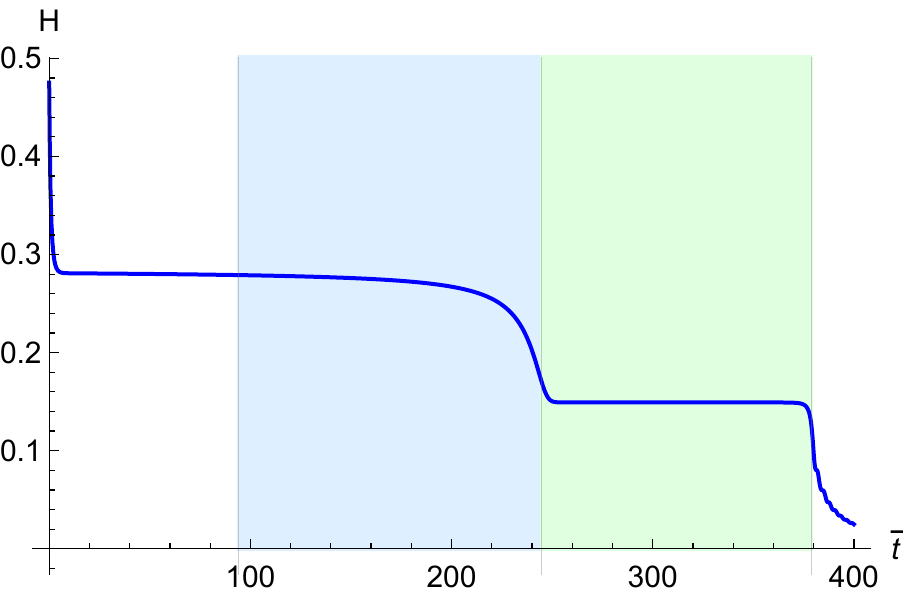}
    \caption{}
    \label{fig_gamma_delta_H}
  \end{subfigure}
  \begin{subfigure}{.32\textwidth}
  \centering
  \includegraphics[width=1\linewidth]{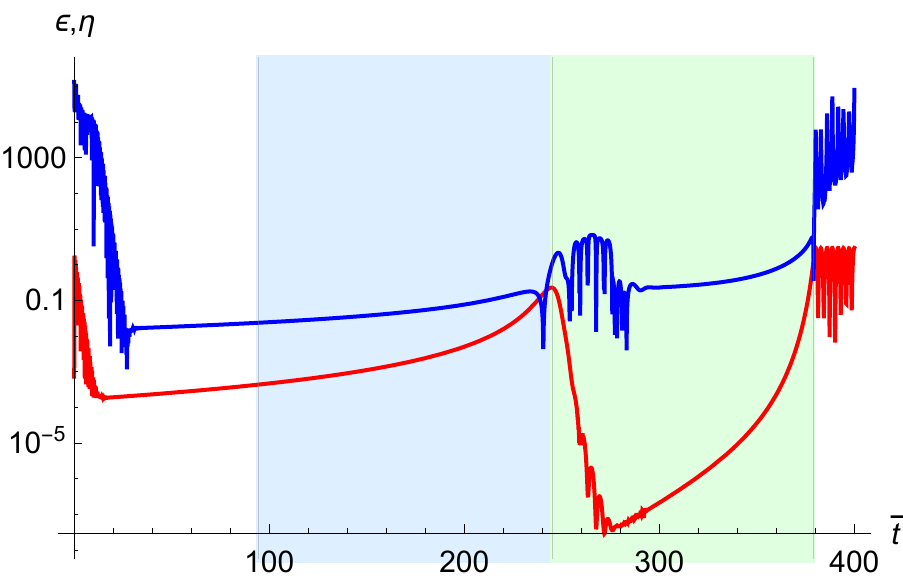}
  \captionsetup{width=.9\linewidth}
  \caption{}
  \label{fig_gamma_delta_para}
  \end{subfigure}
  \captionsetup{width=.9\linewidth}
  \caption{
    (a) The scalar potential of the model, and the two solutions corresponding to the initial conditions $\sigma(0)=0.1$ (blue) and $\sigma(0)=-0.1$ (red), respectively. The  red solution does not have double inflation and should be discarded. 
    (b) The power spectrum around the pivot scale.
    (c) The solution to the field equations \eqref{KG1_gamma} and \eqref{KG2_gamma} for $\varphi$ (red) and $\sigma$ (blue) with the initial conditions $\varphi(0)=6,\sigma(0)=0.1$, the vanishing initial velocities, and the parameters $\delta=0.2$,   $\gamma=10$ and $\zeta=-9.15915$.
    (d) The Hubble function. (e) The SR parameters $\epsilon$ (red) and $\eta$ (blue).
    The blue shaded region represents the first stage of inflation, and the green shaded region represents the second stage.    }
  \label{fig_gamma_delta}
  \end{figure}

 Our results for the CMB observables $n_s$ and $r$ at some values of $\gamma\geq 10$ with $\delta=0.2$, all leading to the sufficient enhancement of the power spectrum for the efficient PBH production, are collected in Table~\ref{tab_gamma_delta1} together with the masses of the induced PBH. 
  
  \begin{table}[ht]
  \centering
  \begin{tabular}{l r r r r r r}
  \toprule
  $\gamma$ & $\zeta$ & $\Delta N_2$ & $n_s$ & $r$ & $M_{\mathrm PBH}$ & $P_{enh.}$ \\
  \hline
  $10$ & $-9.15915$ & $20.00$ & $0.9407$ & $0.0097$ & $3.5\times 10^{16}$ & $3.6\times 10^6$ \\
  $15$ & $-12.06468$ & $19.94$ & $0.9409$ & $0.0096$ & $3.8\times 10^{16}$ & $5.0\times 10^6$ \\
  $20$ & $-14.66315$ & $20.02$ & $0.9407$ & $0.0097$ & $4.9\times 10^{16}$ & $5.1\times 10^6$ \\
  \bottomrule
  \end{tabular}
  \captionsetup{width=.9\linewidth}
  \caption{The values of $n_s$ and $r$ together with the PBH masses $M_{\mathrm{PBH}}$ and the corresponding enhancement $P_{\rm enh.}$ of the scalar power spectrum for some values of $\gamma$ 
  above $10$, with  $\Delta N_2 \approx 20$.}
  \label{tab_gamma_delta1}
\end{table}

It follows from Table~\ref{tab_gamma_delta1} that the scalar tilt $n_s$ is still in tension with the Planck measurements (by $3\sigma$ or more), despite the presence of an extra (adjustable) parameter. Therefore, the naive generalization 
of the special models by using two non-vanishing parameters $\gamma$ and $\delta$ within the Ansatz
defined by Eqs.~(\ref{N_choice2}) and (\ref{N_F_choice2}) does not lead to a perfect match (within $1\s$) with Planck observations, contrary to the expectations raised in Refs.~\cite{Aldabergenov:2020bpt,Aldabergenov:2020yok}. It implies that the "pure" $\delta$-models of inflation and PBH formation, defined in the preceding Section, are special when compared to their hybrid extensions studied in this Subsection, as regards PBH production (as DM) and observational constraints. Those $\delta$-models are distinguished by a {\it chiral} deformation of the modified supergravity potentials, so that we focus on generalizations of the ${\cal F}$ term in the next Subsection.

\subsection{The new models with higher powers of ${\cal R}$ in the ${\cal F}$ term} 
\label{change_term}

Let us now revise our Ansatz (\ref{L_master})  by changing the chiral potential \eqref{N_F_choice2}, because a more general non-chiral potential \eqref{N_choice2} apparently does not lead to a significant improvement in the value of $n_s$, according to our studies in the preceding Subsection.

The linear term in the function ${\cal F}$ is required to generate the Einstein-Hilbert term for gravity, so that 
a generic chiral potential reads
\begin{eqnarray} 
  {\cal F}({\cal R})=-3{\cal R}+f({\cal R})~,
  \label{general_F}
\end{eqnarray}
where the (complex) function $f$ of the chiral superfield ${\cal R}$ should be quadratic at least (we assume it in the polynomial form). 

The simplest new models are defined by using the function $f$ as a higher power $(n>2)$ of ${\cal R}$, 
\begin{eqnarray}
  f({\cal R})\equiv \fracmm{C_n}{M^{n-1}}{\cal R}^n \quad  \mathrm{and}\quad  X=\fracmm{M}{\sqrt{24}}\sigma~,
  \label{new_f}
\end{eqnarray}
where we have ignored the angular mode of $X=\left. {\cal R}\right|$ again.

The effective action for two scalars takes the same form \eqref{L_varphi2}, though with new functions $A,B$ and $U$ 
of $\s$. After a straightforward calculation, we find
\begin{align} 
    A&=1+\fracmm{1}{6}\sigma^2-\fracmm{11}{24}\zeta\sigma^4-\fracmm{29}{54}\gamma\sigma^6-\fracmm{1}{6}(f_X+\overbar{f}_{\overbar{X}})~,\nonumber\\
    B&=\fracmm{1}{3M^2}(1-\zeta\sigma^2-\gamma\sigma^4)~,\label{ABU_tilde}\\
    U&=\fracmm{M^2}{2}\sigma^2\left[1-\fracmm{1}{6}\sigma^2+\fracmm{3}{8}\zeta\sigma^4+\fracmm{25}{54}\gamma\sigma^6
    +\fracmm{1}{3}(f_X+\overbar{f}_{\overbar{X}})-\fracmm{1}{2}\left(\fracmm{f}{X}+\fracmm{\overbar{f}}{\overbar{X}}\right)\right]~.\nonumber 
\end{align}
The $(\gamma,\delta)$ models are reproduced by choosing $f(X)=\frac{3\sqrt{6}}{M}\d X$. 

\subsubsection{The model with a cubic chiral term}

The model is defined by the cubic chiral function $f$ with the new parameter $\lambda$ as follows:
\begin{eqnarray}
  {\cal F}=-3{\cal R}+\fracmm{24}{M^2}\lambda {\cal R}^3~~. \label{n3}
\end{eqnarray} 
The corresponding functions $A, B$ and $U$ are given by
\begin{align} \label{n3f3}
  &A=1+\left( \fracmm{1}{6}-\lambda\right)\sigma^2-\fracmm{11}{24}\zeta \sigma^4-\fracmm{29}{54}\gamma\sigma^6,
  \nonumber \\
  &B=\fracmm{1}{3M^2}\left( 1 -\zeta\sigma^2-\gamma\sigma^4\right), \\
  &U=\fracmm{1}{2}M^2\sigma^2\left[ 1 + \left(\lambda-\fracmm{1}{6}\right)\sigma^2+\fracmm{3}{8}\zeta\sigma^4+\fracmm{25}{54}\gamma\sigma^6 \right]~.\nonumber
\end{align}

The proper values of the parameters $\gamma, \lambda$ and $\zeta$, which are needed for efficient PBH production, 
are collected in Table \ref{tab_gamma_kappa}, where the parameter $\zeta$ has been tuned to get 
$\Delta N_2\approx 20$.

\begin{table}[ht]
  \centering
  \begin{tabular}{l r r r r r}
  \toprule
  $\gamma$ & $1.0$ & $1.0$ & $10$ & $10$ & $20$ \\
  $\lambda$ & $0.1$ & $0.2$ & $0.1$ & $0.2$ & $0.2$ \\
  $\zeta$ & $-1.92564$ & $-2.06496$ & $-9.20329$ & $-9.48826$ & $-15.044427$ \\
  \midrule
  $\Delta N_2$ & $19.97$ & $19.99$ & $20.09$ & $19.95$ & $20.01$ \\
  $n_s$ & $0.9407$ & $0.9402$ & $0.9407$ & $0.9408$ & $0.9408$\\
  $r$ & $0.0097$ & $0.0099$ & $0.0097$ & $0.0096$ & $0.0096$\\
  ${\cal P}_{\mathrm enh.}$ & $3.0\times 10^5$ & $2.1\times 10^6$ & $3.1\times 10^6$ & $4.0\times 10^6$ & $3.8\times 10^6$\\
  ${M}_{\mathrm{PBH}}({\rm g})$ & $1.6\times 10^{16}$ & $1.5\times 10^{11}$ & $4.2\times 10^{16}$ & $3.7\times 10^{16}$ & $5.6\times 10^{16}$\\
  \bottomrule
  \end{tabular}
  \captionsetup{width=.9\linewidth}
  \caption{The parameters $\gamma$, $\lambda$ and $\zeta$ in the cubic model, and the resulting PBH values.}
  \label{tab_gamma_kappa}
  \end{table}

As can be seen from Table~\ref{tab_gamma_kappa}, $\gamma\geq 10$ for $\lambda\sim 0.1$ is required to get a sufficient enhancement of the power spectrum. The value of $n_s$ is sensitive to large changes in the value of 
$\Delta N_2$ that is one of the major factors contributing to the mass $M_{\rm{PBH}}$, as long as the value of $\lambda$ is much smaller than $\gamma$.  The obtained values of $n_s$ are below $0.946$, i.e.  outside the observational constrains by $3\sigma$ or more.

\begin{figure}
  \centering
  \begin{subfigure}{.49\textwidth}
    \centering
    \includegraphics[width=.8\linewidth]{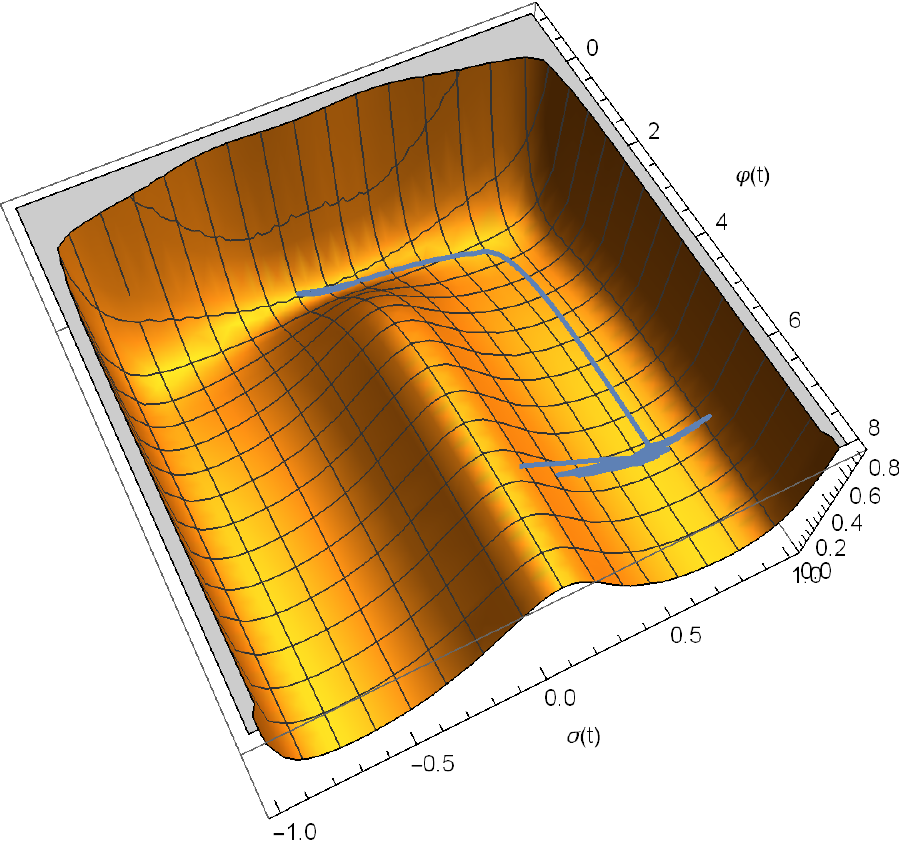}
    \caption{}
    \label{fig_gamma_kappa_V}
  \end{subfigure}
    \begin{subfigure}{.49\textwidth}
    \centering
    \includegraphics[width=.95\linewidth]{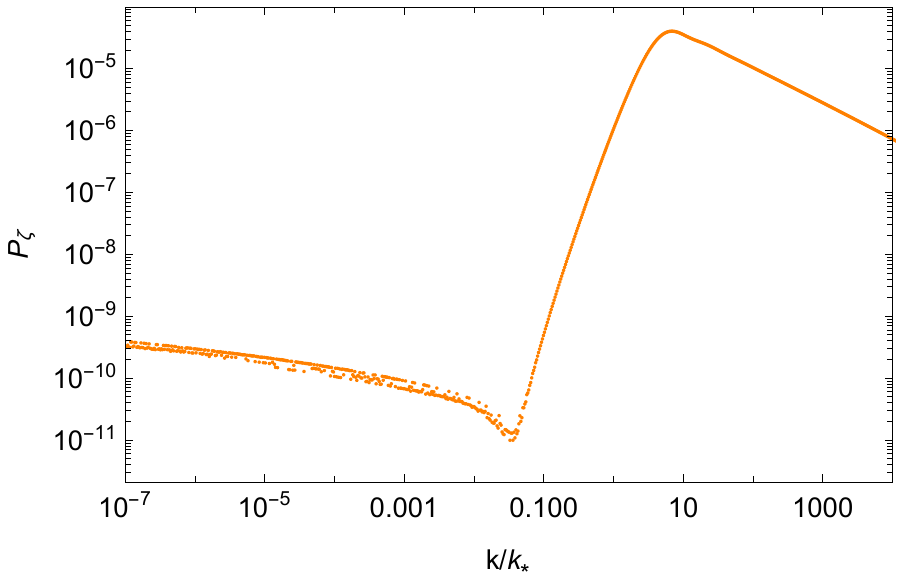}
    \caption{}
    \label{fig_gamma_kappa_pow}
  \end{subfigure}
  \begin{subfigure}{.32\textwidth}
    \centering
    \includegraphics[width=1\linewidth]{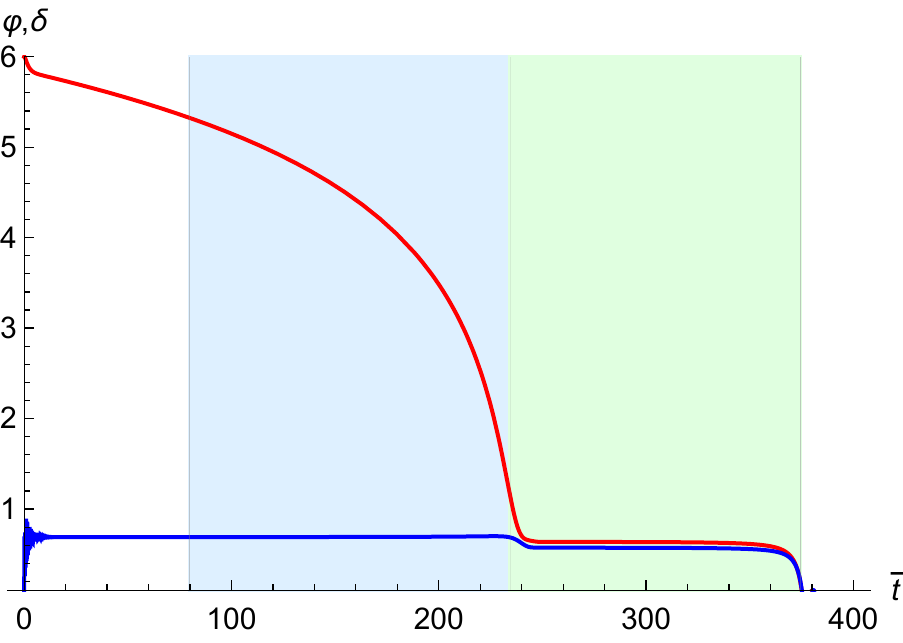}
    \caption{}
    \label{fig_gamma_kappa_sol}
  \end{subfigure}
  \begin{subfigure}{.32\textwidth}
    \centering
    \includegraphics[width=1\linewidth]{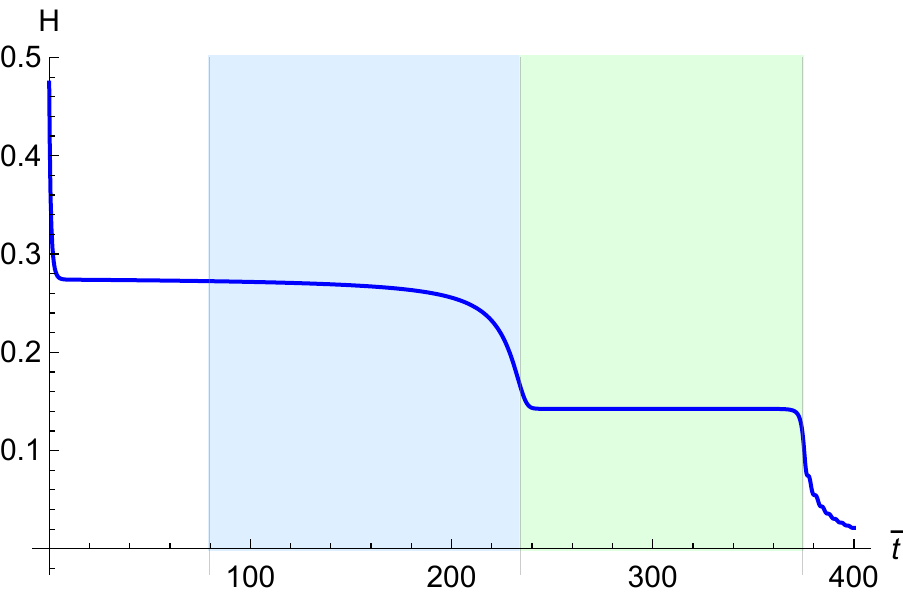}
    \caption{}
    \label{fig_gamma_kappa_H}
  \end{subfigure}  
  \begin{subfigure}{.32\textwidth}
  \centering
  \includegraphics[width=1\linewidth]{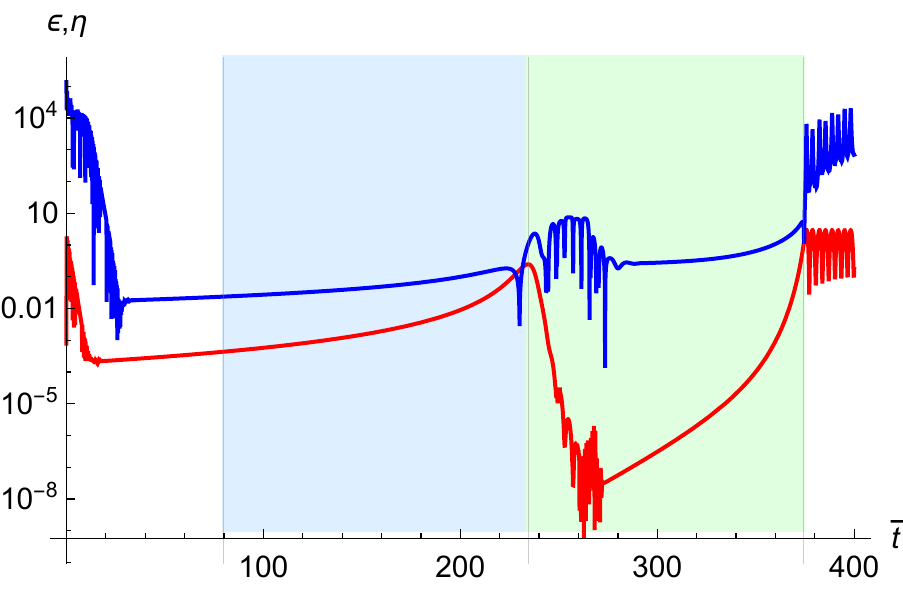}
  \caption{}
  \label{fig_gamma_kappa_para}
  \end{subfigure}

  \caption{
    (a) The shape of the potential and the inflation trajectory in the cubic model with the parameters and the initial conditions given under  (c).
    (b) The power spectrum around the pivot scale.
    (c) The solution to the field equations \eqref{KG1_gamma} and \eqref{KG2_gamma} in the cubic model (\ref{n3}) with the initial conditions $\varphi(0)=6,\sigma(0)=0.1$, the vanishing initial velocities, and the parameters $\lambda=0.2$,   $\gamma=10$ and $\zeta=-9.48826$.
    (d) The Hubble function. (e) The SR parameters $\epsilon$ (red) and $\eta$ (blue) in the cubic model.
    The blue shaded region represents the first stage of inflation, and the green shaded region represents the second stage. }
  \label{fig_gamma_kappa}
  \end{figure}

As a representative of those models, let us take the one with $\gamma=10$, $\lambda=0.2$ and $\zeta=-9.48826$. 
These values are determined by fixing the value of the power spectrum enhancement as $10^6$, after setting the length of the second stage of inflation by about 20 e-folds. Fig.~\ref{fig_gamma_kappa} shows the shape of the potential and the inflationary trajectory (a), the power spectrum (b), the evolution of both scalar fields (c), the shape of the Hubble function (d), and the slow roll parameters (e), in the cubic model with the parameters chosen above.

Unlike the quadratic model,  the potential becomes completely symmetric with respect to a change of the sign of $\sigma$.  As a result of that, the shape of the trajectory is not affected by the sign of the initial condition on $\sigma$. 
It implies that the inflationary trajectory is more stable against changes in the initial value of the inflaton field. 

Using the higher powers $n$ of ${\cal R}$ until $n=7$ in the chiral function ${\cal F}$ leads to the similar results. For instance, the PBH production in the quartic model with $n=4$ appears to be similar to that in the quadratic $\delta$-models, with the potential that is narrower along the $\sigma$-axis and resembles the potentials in the $(\gamma,\delta)$ models with a higher value of $\gamma$. It may have been expected because all those models merely change some coefficients in the functions $A,B$ and $U$. In particular, the value of $n_s$ is not significantly affected.

\subsubsection{The model with $n=9$}

One may wonder, what happens when the power of ${\cal R}$ in the holomorphic potential ${\cal F}$ is increased further in Eq.~(\ref{n3}) ? As is clear from Eq.~(\ref{n3f3}), until $n=9$ it would merely affect some
of the coefficients in the polynomials there, without producing new terms. Therefore, we should not expect significant
changes from the simplest $\delta$-models until $n=8$. Going beyond $n=8$ leads to the appearance of the higher powers of $\s$ according to Eq.~(\ref{ABU_tilde}), when keeping the symmetry under the sign change of  $\sigma$. 

The superpotential ${\cal F}$  for the $n=9$  model with the new parameter $\mu$ reads as follows:
\begin{eqnarray} \label{n9f}
  {\cal F}=-3{\cal R}+\fracmm{(24)^4}{M^8}\mu {\cal R}^9~~.
\end{eqnarray} 

\begin{figure}
  \centering
  \begin{subfigure}{.49\textwidth}
    \centering
    \includegraphics[width=.8\linewidth]{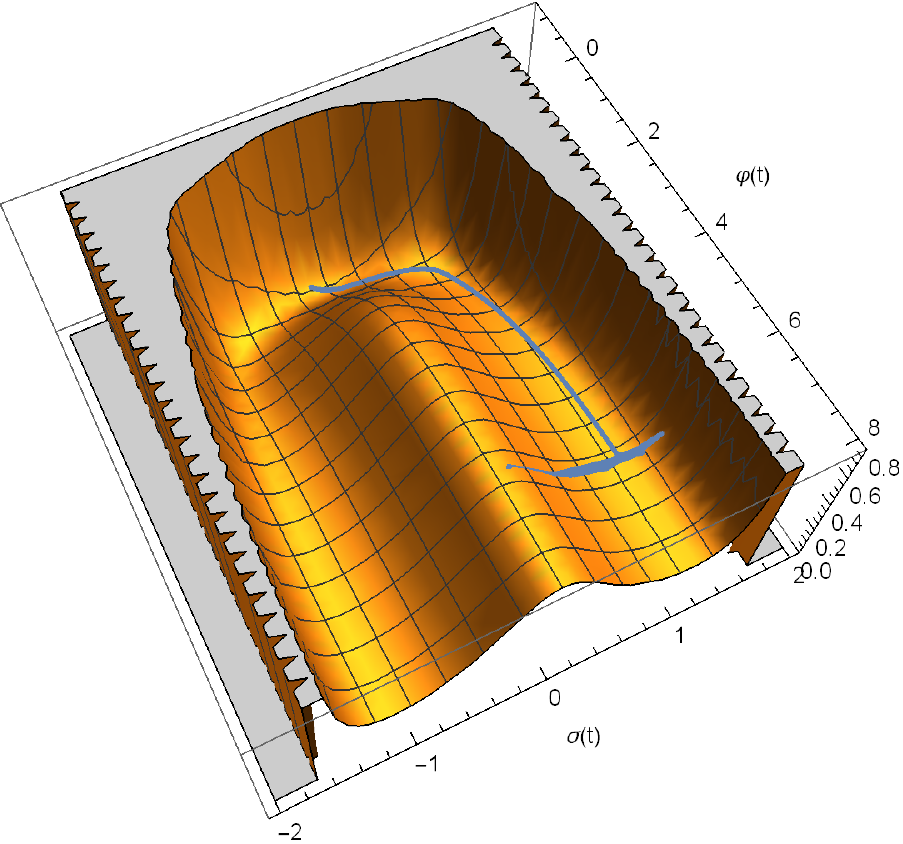}
    \caption{}
    \label{fig_gamma_mu_V}
  \end{subfigure}
  \begin{subfigure}{.49\textwidth}
  \centering
  \includegraphics[width=.95\linewidth]{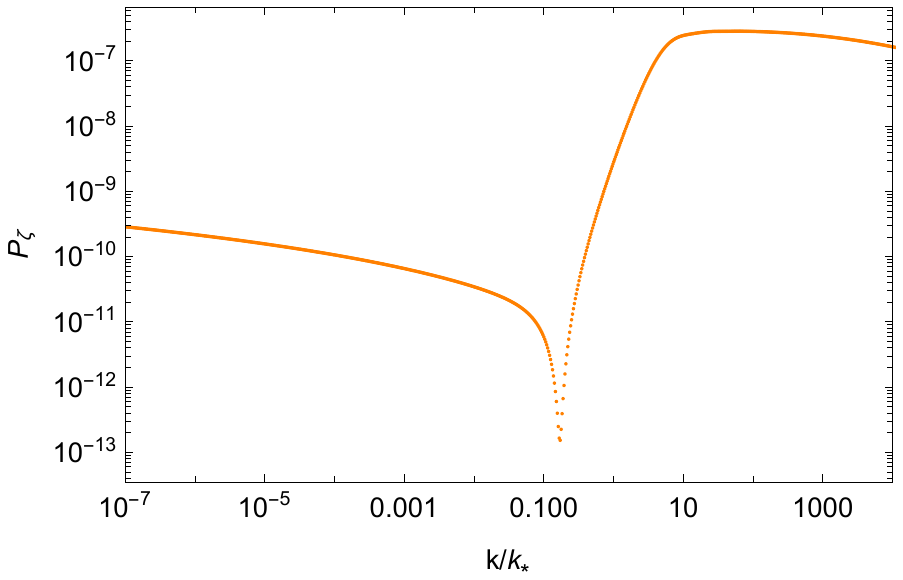}
  \captionsetup{width=.9\linewidth}
  \caption{}
  \label{fig_gamma_mu_pow}
  \end{subfigure}
  \begin{subfigure}{.32\textwidth}
    \centering
    \includegraphics[width=1\linewidth]{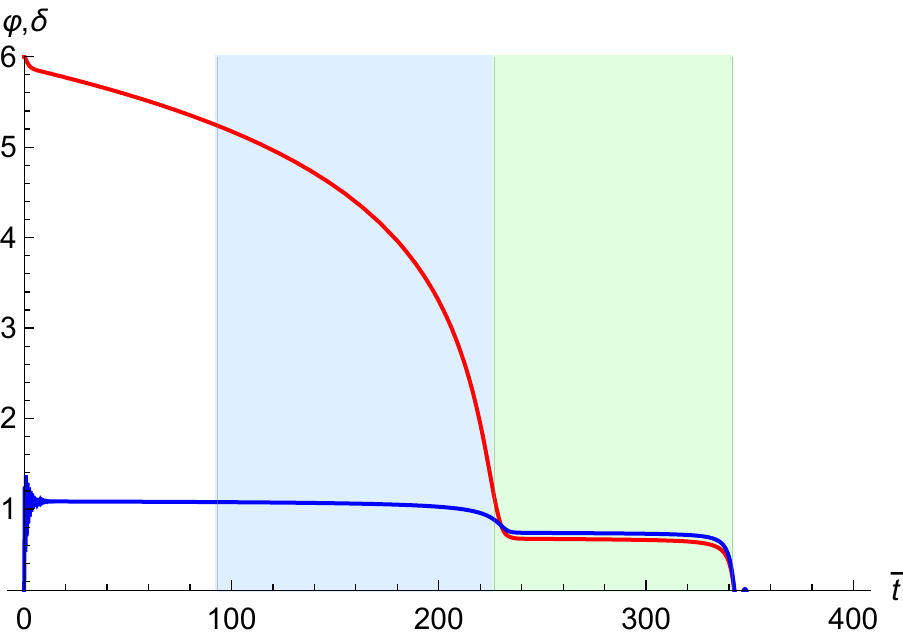}
    \caption{}
    \label{fig_gamma_mu_sol}
  \end{subfigure}
  \begin{subfigure}{.32\textwidth}
    \centering
    \includegraphics[width=1\linewidth]{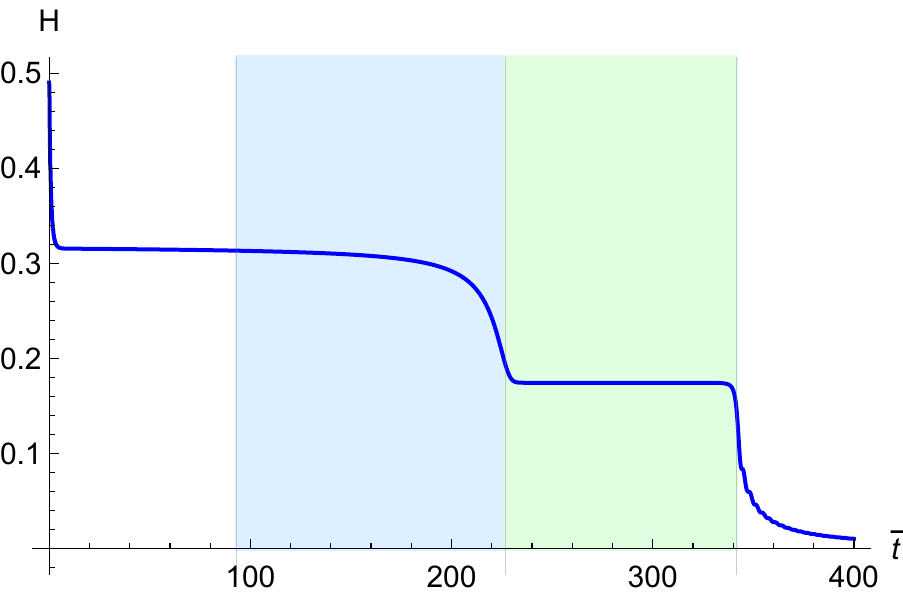}
    \caption{}
    \label{fig_gamma_mu_H}
  \end{subfigure}
  \begin{subfigure}{.32\textwidth}
    \centering
    \includegraphics[width=1\linewidth]{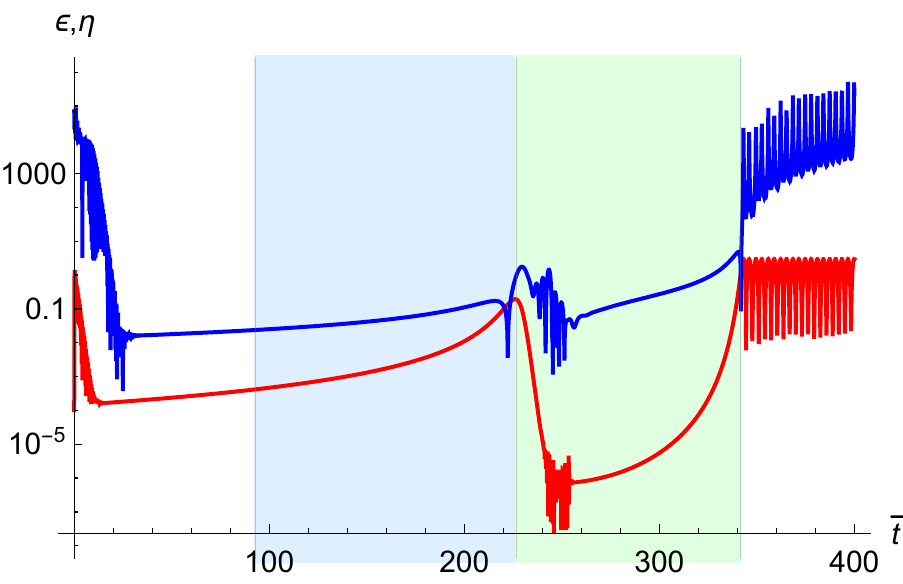}
    \caption{}
    \label{fig_gamma_mu_para}
  \end{subfigure}

  \captionsetup{width=.9\linewidth}
  \caption{
  (a) The scalar potential of the model, and the inflationary trajectory with the initial conditions given under  (c). 
  (b) The power spectrum around the pivot scale.   
  (c) The solution to the field equations \eqref{KG1_gamma} and \eqref{KG2_gamma} in the $n=9$ model (\ref{n9f}) for $\varphi$ 
  (red) and $\sigma$ (blue) with the initial conditions $\varphi(0)=6,\sigma(0)=0.1$, the vanishing initial velocities, 
  and the parameters $\gamma=1.0$,   $\delta=0.1$ and $\zeta=-2.41238$. 
  (d) The Hubble function. 
  (e) The SR parameters, $\epsilon$ (red) and $\eta$ (blue).
  The blue shaded region represents the first stage of inflation, and the green shaded region represents the second stage.
  }
  \label{fig_gamma_mu}
  \end{figure}

As a representative of these models,  let us take $\gamma=1.0$, $\lambda=0.1$ and $\zeta=-2.41238$, towards 
the desired value of $\Delta N_{ 2}\approx 20$ with at least the order $10^6$ enhancement on the scalar power spectrum. The corresponding plots are given in Fig.~\ref{fig_gamma_mu}.
  
After following the procedure described in the preceding Subsection, we find the values collected in Table \ref{tab_n9}.

\begin{table}[ht]
  \centering
  \begin{tabular}{l r r r r r}
  \toprule
  $\gamma$ & $1.0$ & $1.0$ & $1.0$ & $2.0$ & $5.0$ \\
  $\mu$ & $0.01$ & $0.1$ & $0.2$ & $0.1$ & $0.1$ \\
  $\zeta$ & $-1.858261$ & $-2.41238$ & $-2.87306$ & $-3.37249$ & $-5.8009472$ \\
  \midrule
  $\Delta N_2$ & $20.00$ & $20.07$ & $20.06$ & $19.99$ & $20.05$ \\
  $n_s$ & $0.9408$ & $0.9402$ & $0.9314$ & $0.9405$ & $0.9407$\\
  $r$ & $0.0096$ & $0.0093$ & $0.0059$ & $0.0098$ & $0.0097$\\
  ${\cal P}_{\mathrm enh.}$ & $4.3\times 10^6$ & $1.9\times 10^6$ & $3.7\times 10^6$ & $4.1\times 10^6$ & $2.7\times 10^6$\\
  ${M}_{\mathrm{PBH}}({\rm g})$ & $3.3\times 10^{16}$ & $6.6\times 10^{14}$ & $7.0\times 10^{11}$ & $3.2\times 10^{16}$ & $5.6\times 10^{16}$\\
  \bottomrule
  \end{tabular}
  \captionsetup{width=.9\linewidth}
  \caption{The parameters $\gamma$, $\mu$ and $\zeta$ in the $n=9$ model (\ref{n9f}), and the resulting PBH values.}
  \label{tab_n9}
  \end{table}

As is clear from Table \ref{tab_n9},  the enhancement of the scalar power spectrum in the $n=9$ model (\ref{n9f}) 
can be as large as $10^6$ with the relatively smaller values of $\gamma$  in comparison to the previous models, but the predicted values of the tilt $n_s$ are once again outside the $3\sigma$ region of the Planck measurements. The calculated value of $n_s$ increases when the value of $\mu$ decreases, which leads us back to the 
$\gamma$-models. It appears to be difficult to increase the masses of the generated PBH  beyond $10^{16}$ g in the 
$n=9$ model. The parameter $\zeta$ needs to be fine-tuned as before.

Therefore, it seems to be impossible to improve the values of cosmological observables (mainly $n_s$) by changing the second term in Eq.~\eqref{N_F_choice2} to another term with a higher power.
 
\section{Conclusion}

Our extensions of the Starobinsky inflation model in the modified supergravity are capable to produce PBH that may account for part or the whole Cold Dark Matter, with the PBH masses in the range between $10^{16}$ g and $10^{20}$ g. For example, the NANOGrav Collaboration data \cite{Arzoumanian:2020vkk} hints to the PBH as DM \cite{DeLuca:2020agl}, in agreement with our results in Fig.~\ref{Fig_Pf}.

The PBH generation can be efficiently catalyzed by primordial perturbations sourced by Starobinsky's scalaron coupled to another scalar of the (super)gravitational origin, leading to the effective two-field inflation. Our conclusions are based on theoretical considerations with the two fundamental principles: modified gravity and supersymmetry, starting from the manifestly (locally) supersymmetric Lagrangians, in the top-down approach. 

Our models have a limited number of free parameters that have to be properly tuned. Surprisingly enough, a naive increase in the number of the parameters does not relax their tuning needed for viable inflation and efficient PBH production (as DM). We found that it is highly non-trivial (and, perhaps, impossible) to meet the $1\sigma$ Planck value of $n_s$ in our models for a large part of the parameter space, though it is still possible to get an agreement between $2\sigma$ and $3\sigma$ in the $\delta$-models with $\gamma=\lambda=\mu=0$,

It is worth mentioning that we insisted on the duration of the second stage of inflation to be close to 20 e-folds that allows the whole DM as PBH. Should the lower $n_s$ values (close to $0.946$) be excluded by $5\sigma$,  only a small part  (under 10\%) of DM as PBH can be realized in our models.

Isocurvature perturbations play the key role for PBH formation in our approach because their growth during USR is part of our amplification mechanism. Large isocurvature perturbations appear when the effective isocurvature mass becomes tachyonic and the inflationary trajectory makes a sharp turn. It is the genuine multi-field inflation effect that cannot be realized in single-field models of inflation,  see e.g., Refs.~\cite{Ketov:2021fww,Gundhi:2020kzm} for details. Our mechanism of the amplification of curvature perturbations is different from the one employed in Ref.~\cite{Braglia:2020eai} where the kinetic coupling between two scalars is responsible for the tachyonic isocurvature mass.

As is well known, PBH formation necessarily leads to gravitational waves (GW) because large scalar over-densities act as a source of the stochastic GW background.  Frequencies of those GW can be related to the expected PBH masses and the duration of the second stage of inflation \cite{Cai:2018dig,Bartolo:2018evs}. Those GW may be detected in the future  ground-based experiments, such as the Einstein telescope \cite{ET} and the global network of GW interferometers including the advanced LIGO, Virgo and KAGRA \cite{ALAVK}, as well as the space-based GW interferometers such as LISA \cite{LISA}, TAIJI (old ALIA) \cite{TAIJI,Ruan_2020}, TianQin \cite{TQ} and DECIGO \cite{DEC}.

Our models of inflation and PBH production may be tested by observations via detection of the stochastic GW induced by the PBH formation.  Using the power spectra on the left side of Fig.~\ref{Fig_Pf}, the density $\Omega_{\rm GW}(k)$ in terms of frequency $k=2\pi f$ was computed in Ref.~\cite{Aldabergenov:2020yok}, with the results collected in Fig.~\ref{Fig_Omega_GW} including  the expected sensitivity curves for several space-based GW experiments planned in the near future.  To draw the sensitivity curves, the parameters and the noise models for LISA \cite{LISA},  TianQin \cite{TQ}, Taiji \cite{TAIJI}, and DECIGO \cite{DEC} have been used. Our Fig.~\ref{Fig_Omega_GW} here is different from the one given in Ref.~\cite{Aldabergenov:2020yok} because the updated sensitivity curve of the Taiji project \cite{Ruan_2020} has been used.

It follows from Fig.~\ref{Fig_Omega_GW} that the upcoming space-based GW experiments may be sensitive enough to detect the stochastic GW background predicted by our models. According to Fig.~\ref{Fig_Omega_GW}, our supergravity models produce GW peaks in the frequency range $10^{-3}\div 10^{-1}$ Hz expected to be accessible by the LISA, TianQin, Taiji, and DECIGO gravitational interferometers.

It is yet to be seen to what extent future observations of CMB and GW would allow us to reconstruct our models and distinguish them from others. One possibility is to look at spectral distortions as a probe of primordial density perturbations, as is proposed in Ref.~\cite{2021ExA...tmp...42C}. Dissipation of those perturbations through photon
diffusion will distort the CMB spectrum at an observable level \cite{Silk:1967kq,1970Ap&SS...9..368S}. Another possibility may be looking at specific oscillatory features of the stochastic GW background~\cite{Braglia:2020taf}.

\begin{figure}
\includegraphics[width=0.8\linewidth]{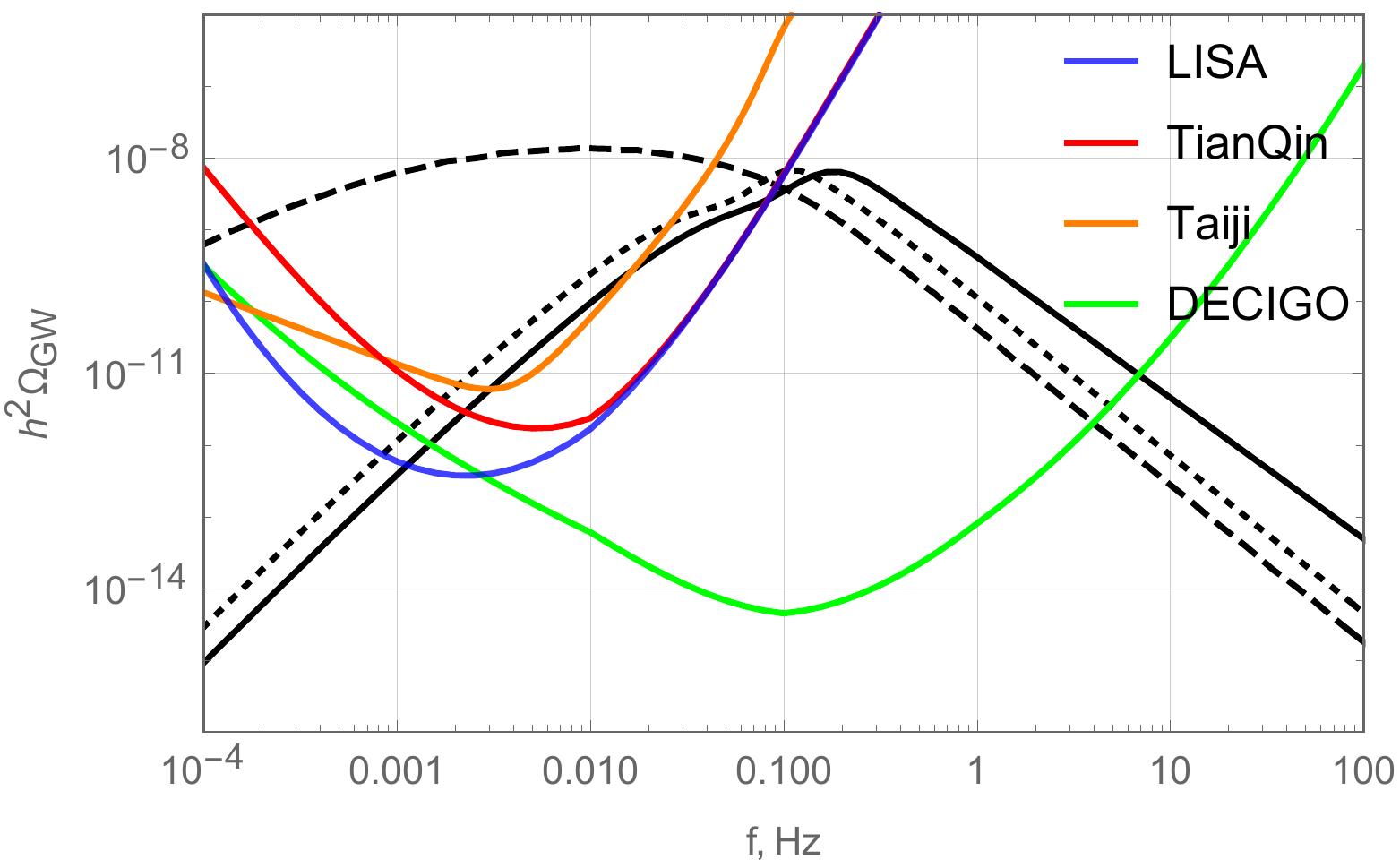}
\caption{\label{Fig_Omega_GW} The density of stochastic GW induced by the power spectrum enhancement in the supergravity models studied in Subsection 3.3: the case I (solid black curve), the case II (dashed black curve), and the case III (dotted black curve). The expected sensitivity curves for the space-based GW experiments are represented by the different colors.}
\end{figure}

Fine tuning in our best models amounts to fixing the parameter $M\sim 10^{-5}M_{\rm Pl}$ as the scalaron mass,  the dimensionless parameter $\zeta$ for the desired duration of the USR phase between two stages of inflation, and the parameter $\delta$ for agreement with the Planck measurements of $n_s$. The obtained PBH masses and cosmological tilts agree with all astrophysical and cosmological constrains, including $n_s$ (in the $\delta $-models). We are not aware of any fundamental reason for the $\delta$-models to be selected, beyond the phenomenological constraints.

Supergravity is usually regarded as a fundamental extension of gravity at super-high energy scales. Our new findings demonstrate that the new scalars of modified supergravity can play the important role during inflation, catalyze PBH formation and produce GW radiation. The interactions of those scalars are dictated by local SUSY and are {\it not}  assumed {\it ad hoc}. Our models thus have the predictive power that can be confirmed or falsified in future experiments, also implying that indirect footprints of SUSY may be detectable from GW physics rather than high-energy particle colliders!

\newpage

\section*{Acknowledgements}

RI and SVK were supported by Tokyo Metropolitan University. SVK was also supported by the World Premier International Research Center Initiative, MEXT, Japan, and the Development  Program of Tomsk Polytechnic University within the assignment of the Ministry of Science and Higher Education of the Russian Federation.
The authors are grateful to M. Braglia, J. Chluba, H. Murayama, S. Pi and A.A. Starobinsky for discussions and correspondence.

\bibliography{Bibliography}{}
\bibliographystyle{utphys}

\end{document}